\RequirePackage{fix-cm}
\documentclass[twocolumn]{svjour3}          
\smartqed  
\usepackage{graphicx}
\usepackage{eqnarray}
\usepackage{amsmath}
\usepackage{amssymb}
\usepackage[round]{natbib}
\usepackage{bibentry}
\usepackage{nameref}
\usepackage{color}
\usepackage[normalem]{ulem}
\usepackage{xcolor}
\begin{document}

\title{Dynamics of spontaneous activity in random networks with multiple neuron subtypes and synaptic noise
\thanks{This paper was developed within the scope of the IRTG 1740 / TRP 2015/50122-0, funded by DFG / FAPESP. RP and ACR are also part of the Research, Innovation and Dissemination Center for Neuromathematics (FAPESP grant 2013/07699-0). RP is supported by a FAPESP scholarship (2013/25667-8). ACR is partially supported by a CNPq fellowship (grant 306251/2014-0).}
}

\subtitle{Spontaneous activity in networks with synaptic noise}


\author{Rodrigo F.O. Pena \and 
		Michael A. Zaks$^\dagger$ \and 
        Antonio C. Roque$^\dagger$
}

\authorrunning{Pena et al.} 

\institute{Rodrigo F.O. Pena \at
              Dept of Physics, Faculty of Philosophy, Sciences and Letters of Ribeir\~{a}o Preto, University of S\~{a}o Paulo, Ribeir\~{a}o Preto, SP, Brazil \\
              \email{rfdop@uol.com.br}           
           \and
           Michael A. Zaks \at
              Dept of Physics, Faculty of Mathematics and Natural Sciences, Humboldt University of Berlin, Berlin, Germany \\
              \email{zaks@physik.hu-berlin.de} 
              \and 
              Antonio C. Roque \at
               Dept of Physics, Faculty of Philosophy, Sciences and Letters of Ribeir\~{a}o Preto, University of S\~{a}o Paulo, Ribeir\~{a}o Preto, SP, Brazil \\
              \email{antonior@usp.br} 
              \\
              $^\dagger$ These authors contributed equally to this work.
}


\maketitle

\begin{abstract}
Spontaneous cortical population activity exhibits a multitude of oscillatory patterns, which often display synchrony during slow-wave sleep or under certain anesthetics and stay asynchronous during quiet wakefulness. The mechanisms behind these cortical states and transitions among them are not completely understood. Here we study spontaneous population activity patterns in random networks of spiking neurons of mixed types modeled by Izhikevich equations. Neurons are coupled by conductance-based synapses subject to synaptic noise. We localize the population activity patterns on the parameter diagram spanned by the relative inhibitory synaptic strength and the magnitude of synaptic noise. In absence of noise, networks display transient activity patterns, either oscillatory or at constant level. The effect of noise is to turn transient patterns into persistent ones: for weak noise, all activity patterns are asynchronous non-oscillatory independently of synaptic strengths; for stronger noise, patterns have oscillatory and synchrony characteristics that depend on the relative inhibitory synaptic strength. In the region of parameter space where inhibitory synaptic strength  exceeds the excitatory synaptic strength and for moderate noise magnitudes networks feature intermittent switches between oscillatory and quiescent states with characteristics similar to those of synchronous and asynchronous cortical states, respectively. We explain these oscillatory and quiescent patterns by combining a phenomenological global description of the network state with local descriptions of individual neurons in their partial phase spaces.  Our results point to a bridge from events at the molecular scale of synapses to the cellular scale of individual neurons to the collective scale of neuronal populations.
\end{abstract}

\section{ Introduction}
\label{intro}

Simultaneous recordings from large neuronal populations
disclose complex spatio-temporal firing patterns characterized by rhythmic oscillations with variable degrees of synchrony \citep{buzsaki2004,bonifazi2009,uhlhaas2009,colgin2011}. Recent evidence suggests that in the cortex these patterns range from a ``synchronized" state, characterized by low-frequency oscillation in the population firing rate and up/down switching in the single-neuron membrane potential, to a ``desynchronized"  state, marked by a roughly constant population firing rate and irregular single-neuron firing \citep{harris2011,vyazovskiy2011,sachidhanandam2013,miller2014,okun2015,jercog2017}. Synchronous states are more prominent during slow-wave sleep and anesthesia whereas asynchronous firing activity is prevalent in the states of wakefulness and REM sleep \citep{steriade2001,elboustani2007,greenberg2008,sanchez-vives2017}. Notably, the degree of synchrony in cortical and subcortical regions varies with time, often with intermittent switches between synchronous and asynchronous states \citep{rubchinsky2013,hahn2017,rubchinsky2017}.

There is a widespread assumption that prevalence of synchrony or asynchrony in the network activity depends on the relative strength of excitatory and inhibitory synaptic inputs \citep{vanvreeswijk1996,amit1997,renart2010,landau2016}. In the context of networks of leaky integrate-and-fire (LIF) neurons, the balance between average excitatory and inhibitory synaptic inputs is known to result in quantitative characteristics of network activity that resemble those of asynchronous cortical states \citep{brunel2000dynamics,mattia2002,cessac2008,vogels2005,kumar2008,wang2011,litwin-Kumar2012,kriener2014,ostojic2014,potjans2014}. In the absence of such balance, the network displays behaviors akin to synchronous cortical states \citep{vogels2005review}. 

Networks in which the nodes feature more complicated dynamics than LIF neurons and are able to reproduce intrinsic firing patterns of contrasting cortical neurons, e.g. based on the Izhikevich \citep{izhikevich2003simple,izhikevich2007} or the AdEx \citep{brette2005,gerstner2014} models, demonstrate higher diversity of temporal patterns. In the region of parameter space where inhibitory synaptic strength exceeds excitatory synaptic strength,  mixtures of neurons with different individual firing characteristics perform collective spontaneous oscillations that resemble the alternation of up and down states observed in the synchronized cortical state \citep{tomov2014,tomov2016}. This suggests that not only synaptic balance of excitation/inhibition but also heterogeneities in the neuronal composition of the network may have an impact on the dynamic pattern of the network. 

Yet another factor, capable of influencing the interplay between oscillatory and non-oscillatory states, is the intrinsic randomness of synaptic channels. More specifically, stochasticity expressed by synaptic noise originates from spontaneous neurotransmitter release in the synaptic cleft which generates miniature excitatory (inhibitory) postsynaptic potentials, the so-called mEPSPs (mIPSPs) or simply minis \citep{kavalali2015,pulido2017}. Characteristics of miniature postsynaptic potentials
as amplitude and frequency have been demonstrated to depend on the sleep/wake state \citep{liu2010}. From the theoretical point of view, synaptic noise has been used in cortical models as a source of transitions between different dynamical network states \citep{compte2000,renart2003,holcman2006,moreno-bote2007,parga2007}.   

Previous work has shown that up-down oscillations can appear in different setups. One of them considers neurons with an adaptive variable, within e.g. AdEx \citep{destexhe2009} or Izhikevich \citep{tomov2014} formalism. Another setup uses noise to provoke the switching between the two states \citep{holcman2006,jercog2017}. Here, by combining adaptation with noise, we show that noise is not mandatory for the up-down oscillations but favors their occurrence when it is present.   
In this study we demonstrate that a network of Izhikevich neurons with stochastic synaptic inputs displays a rich variety of dynamic states with different levels of oscillations and degrees of synchrony. We locate these states in the parameter space spanned by the ratio between inhibitory and excitatory synaptic increments and the synaptic noise magnitude. As expected, noise transforms the transient dynamics observed in previous studies into persistent states with well established properties. Independently of network composition and relative inhibitory synaptic strength, for low intensities of synaptic noise the persistent states are asynchronous and non-oscillatory. For higher noise magnitudes, the type of persistent state depends on the relative inhibitory synaptic strength. 

Remarkably, in the region of the parameter space where inhibitory synaptic increments are greater than excitatory synaptic increments the persistent state displays intermittent spontaneous transitions between two dynamic regimes: an active state characterized by rhythmic alternations of tonic firing and silence, and a quiescent state characterized by low-rate irregular network firing. In the active state, the average neuronal membrane voltage oscillates between depolarized and hyperpolarized states in a manner that resembles cortical up/down oscillations, whereas in the quiescent state the average membrane potential remains close to the resting value. We characterize this intermittent state by means of firing rates, power spectra, voltage series, and explain the observed phenomena in terms of the behavior of network-embedded neurons viewed in their single-neuron phase subspaces.   

This work extends previous studies on activity pattern dynamics in random networks of LIF neurons to random networks with more involved on-site dynamics.
To test the validity  of our observations against the change of the chosen neuronal model, we
performed similar computations for the same networks composed of the AdEx neurons,
reproducing all basic effects found for the Izhikevich model. This paves way to a broader conjecture that 
two-dimensional neuron models with a slow recovery variable 
can naturally account for oscillations between depolarized and hyperpolarized states, mimicking up/down states. In this context, the synaptic noise can transform transient oscillatory network activity into a persistent complex state with intermittent switches between two different dynamic regimes. 

\section {Materials and methods}

\subsection {Neuron and network model}
\begin{figure}[!ht]
\includegraphics[scale=0.35]{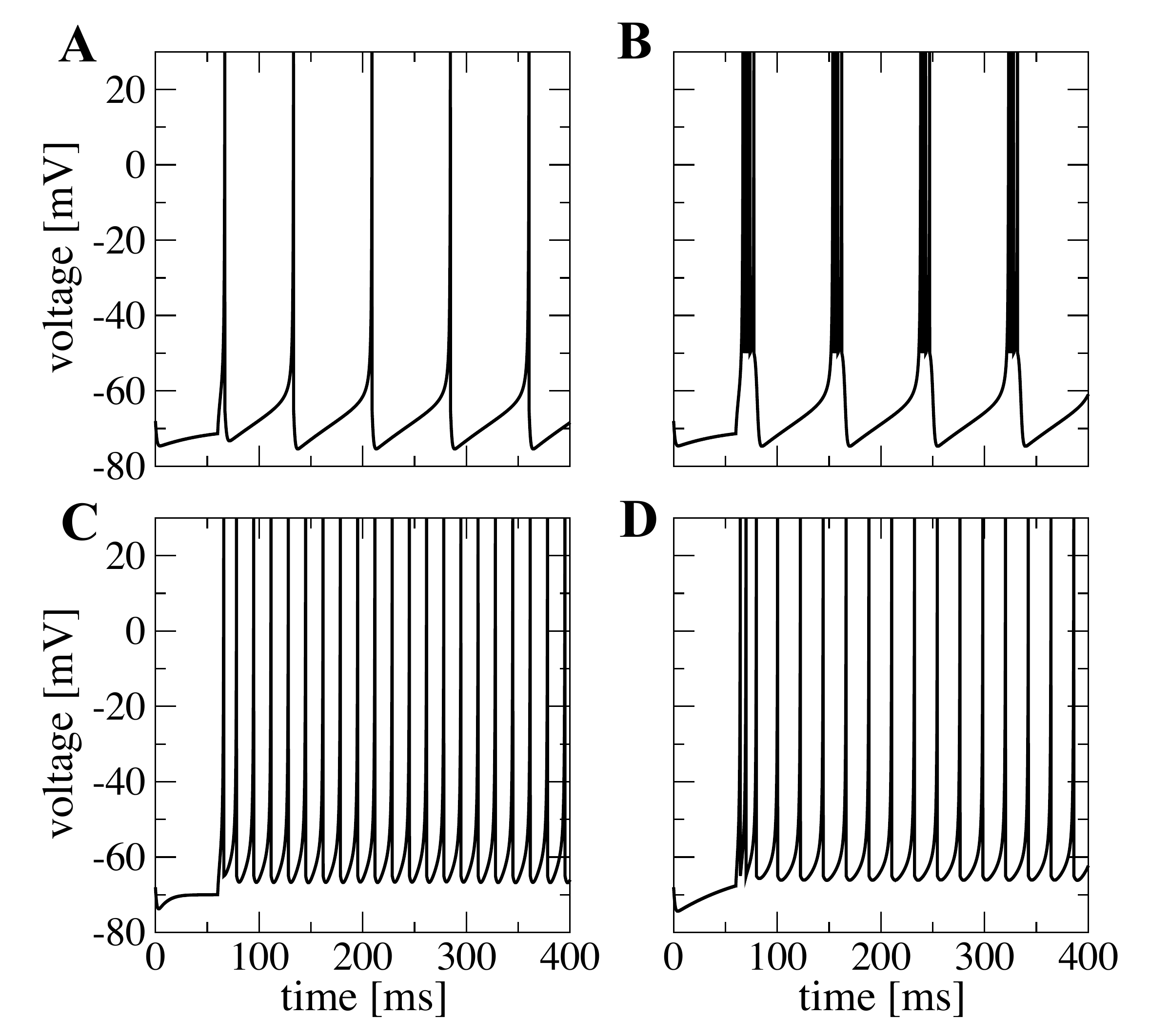}
\caption{\textbf{Spiking patterns for electrophysiological cell classes modeled by the Izhikevich formalism}. \textbf{A} excitatory neuron RS. \textbf{B} excitatory neuron CH. \textbf{C} inhibitory neuron FS. \textbf{D} inhibitory neuron LTS. Plots where produced with constant $I=6$.}
\label{Fig:classes}
\end{figure}

Our work is based on a recent model that describes self-sustained oscillations across high (up) and low (down) global activity states \citep{tomov2014,tomov2016}.
This is the standard random network model where directed connections between every two nodes exist with a fixed probability $p$. To keep cortical sparseness we have chosen a low connection probability $p=0.01$ and size $N=2^{10}$. This renders the expected number of incoming connections per node (average in-degree) $p(N-1)\approx 10$. The network is mixed: it includes both excitatory and inhibitory nodes. The sizes of excitatory and inhibitory subpopulations are taken in the proportion $4:1$ \citep{brunel2000dynamics}.
Each network node is a neuron modeled by the Izhikevich formalism \citep{izhikevich2003simple} with parameters that ensure diverse dynamics on the individual level. Every neuron is described by two variables: voltage $v(t)$ and membrane recovery variable $u(t)$, which follow the coupled differential equations
\begin{eqnarray}
\begin{cases}
\dot{v} & =\alpha v^{2}+\beta v+\gamma - u + I(t) \\
\dot{u} & =a(bv-u), 
\end{cases}
\label{eq:Izh-neuron}
\end{eqnarray}
with a fire and reset rule. Every time when $v(t)$ assumes the threshold value $v(t)=v_{\rm peak}$, both variables are instantaneously updated:
\begin{eqnarray}
\begin{cases}
v(t) & \rightarrow c,\\
u(t) & \rightarrow u(t)+d.
\end{cases}
\label{eq:Izh-updates}
\end{eqnarray}
Our choice of the Izhikevich neuronal model is based on its ability to mimic, by means of setting the appropriate values of parameters $a,b,c,d$, the behavior of neurons from different electrophysiological classes \citep{nowak2003,contreras2004}. Among those, we concentrate in this study on the excitatory regular spiking (RS) and chattering (CH) neurons, and on the inhibitory fast spiking (FS) and low-threshold spiking (LTS) neurons. Fig~\ref{Fig:classes} shows examples of individual dynamics for different classes: the neuronal types differ in frequency, adaptation, and in rheobase current. 
We consider network compositions where all
inhibitory neurons belong to the same class: all of them are either of the LTS type or of the FS type. In the excitatory subpopulation we take the case when all neurons belong to the RS type, and the case when the RS neurons are mixed with CH. A thorough discussion of different aspects of the Izhikevich neuron model can be found in \cite{izhikevich2007}.

The last term in the first equation of (\ref{eq:Izh-neuron}) describes the synaptic current which, for a neuron $j$, reads:
\begin{equation}
I_{syn,j}(t)=G^{\rm ex}_{j}(t)\left(E_{\rm ex}-v_{j}\right) + G^{\rm in}_{j}(t)\left(E_{\rm in}-v_{j}\right).
\label{Eq:current}
\end{equation}
The current is controlled by conductances $G_{j}^{\rm ex/in}$ and reversal potentials 
$E_{\rm ex/in}$, responsible for excitatory/in\-hibitory effects.
Whenever an excitatory (inhibitory) neuron spikes, an increment $g_{\rm ex}$ ($g_{\rm in}$) is added to the conductances $G^{\rm ex}$ ($G^{\rm in}$) of all its postsynaptic neurons; thereafter the conductances decay exponentially with time constant $\tau_{\rm ex/in}$. 
This is well known as a conductance based synaptic model,
described by the differential equation 
\begin{multline}
\frac{dG_{j}^{\rm ex/in}(t)}{dt}=-\frac{G_{j}^{\rm ex/in}(t)}{\tau_{\rm ex/in}}+g_{\rm ex/in} \sum_i \delta(t - t_i) \\
+ \sqrt{2D n_j}\xi_j(t),
\label{Eq:synapse}
\end{multline}
where summation is performed over all time instants $t_i$ of preceding
presynaptic spikes.
We adopt the same parameters as in \cite{tomov2016}: $E_{\rm ex}=0$ mV, $E_{\rm in}=-80$ mV, $\tau_{\rm ex}=5$ ms and $\tau_{\rm in}=6$ ms.

The last term in Eq.(\ref{Eq:synapse}) is the synaptic noise source. Since, for simplification, the noise sources are treated as being independent or weakly correlated, a superposition of a large number of such inputs is approximated by a simple Gaussian white noise process. We assume that $\xi_j$ is Gaussian with zero mean and unit variance: 
$\left\langle \xi(t)\right\rangle =0$ and 
$\left\langle \xi(t)\xi(s)\right\rangle =\delta(t-s)$. Note that in spite of the  zero mean of the Gaussian process, the mean value of the synaptic input current $I_{syn,j}$ stays non-zero which, in its turn, is determined by both $G_j^{\rm ex/in}$ and $E_{\rm ex/in}$. So, the Gaussian process only has the effect of causing displacements in the synaptic current but does not act as a driving force. Concerning the variance, since the sum of independent random normally distributed variables is normally distributed as well, the overall variance of the stochastic
process for a neuron $j$ is chosen to be proportional to the total number of excitatory/inhibitory inputs $n_j$ that this neuron receives. Thereby, for neurons with different numbers of presynaptic partners, the intensity of the noisy input is different. Altogether, evolution of
conductances 
for each neuron consists of the stochastic Ornstein-Uhlenbeck process \citep{uhlenbeck1930} in the time intervals {\em between} the presynaptic spikes and of discontinuous jumps upwards of the size $g_{\rm ex/in}$ at the instants of arrival of those spikes. 
This stochastic model, similar to the point-conductance model described in \cite{destexhe2001}, has its power-spectral density and variance completely determined \citep{gillespie1996}. In distinction to \cite{destexhe2001}, in our case randomness is generated within the synapses and is, in general, non-Poissonian.

The complete set of parameter values used in the simulations of this study is summarized in Table~\ref{Tab::parameters}. Note that that the values of parameters $\alpha, \beta, \gamma$ in the voltage equation are shared by all neuronal types.
\begin{table*}[!ht]
\begin{centering}
\caption{\bf{Parameters used in the simulations.}}
\begin{tabular}{ccccccc}
\textbf{Common parameters in Eq.(\ref{eq:Izh-neuron}) } &  & \textbf{$\textbf{\ensuremath{\alpha}}$} & \textbf{$\textbf{\ensuremath{\beta}}$} & \textbf{$\textbf{\ensuremath{\gamma}}$} & $v_{peak}$ {[}mV{]} & \tabularnewline
\hline 
 &  & 0.04 & 5 & 140 & 30 & \tabularnewline
 &  &  &  &  &  & \tabularnewline
\textbf{Parameters of (\ref{eq:Izh-neuron}) for} &  &  &  &  &  & \tabularnewline
\textbf{different firing patterns } &  & \textbf{a} & \textbf{b} & \textbf{c {[}mV{]}} & \textbf{d} & \tabularnewline
\hline 
\textbf{Excitatory RS} &  & 0.02 & 0.2 & -65 & 8 & \tabularnewline
\textbf{Excitatory CH} &  & 0.02 & 0.2 & -50 & 2 & \tabularnewline
\textbf{Inhibitory} \textbf{FS} &  & 0.1 & 0.2 & -65 & 2 & \tabularnewline
\textbf{Inhibitory LTS} &  & 0.02 & 0.25 & -65 & 2 & \tabularnewline
 &  &  &  &  &  & \tabularnewline
\textbf{Synaptic parameters} & $g_{\rm ex}^{\rm max}$ & $g_{\rm in}^{\rm max}$ & $\tau_{\rm ex}$ {[}ms{]} & $\tau_{\rm in}$ {[}ms{]} & $E_{\rm ex}$ {[}mV{]} & $E_{\rm in}$ {[}mV{]}\tabularnewline
\hline 
 & 0.15 & 1 & 5 & 6 & 0 & -80\tabularnewline
 &  &  &  &  &  & \tabularnewline
\textbf{Network parameters} & size $N$ & ratio exc:inh & connectivity &  &  & \tabularnewline
\hline 
 & $2^{10}$ & 4:1 & $p=0.01$ &  &  & \tabularnewline
\end{tabular}
\label{Tab::parameters}
\par\end{centering}
\end{table*}

\subsection {Measures}
\label{subsec:measures}

In this subsection, we introduce neuron and network measures that will be used below for characterization of the results. 

We start by defining the network time-dependent firing rate as
\begin{equation}
r(t; \Delta t) = \frac{1}{N \Delta t} \sum_{j=1}^{N} \int_{t}^{t+\Delta t} x_j(t') dt',
\label{Eq:firing_rate}
\end{equation}
where the spike train $x_j$ for each neuron $j$ is viewed as a series of $\delta$ functions: $x_j(t) = \sum_{t^f_j} \delta(t-t^f_j)$, with $\{t^f_j\}$ being the set of times when neuron $j$ fired. We fix the time window $\Delta t = 1$ ms.

We will use two power spectra: the spike train power spectrum and the voltage time series power spectrum. The first one is defined for each neuron $j$ as

\begin{equation}
S_{xx,j}(f) = \frac{\langle\tilde{x}_j\tilde{x}^{*}_j\rangle}{T},
\label{Eq:spk_train_spectrum}
\end{equation}
where $\tilde{x}_j(f)$ is the Fourier transform $\tilde{x}_j(f) = \\ \int_{0}^{T}dt\,e^{2\pi ift}x_j(t)$, $\tilde{x}_j^*$ is the complex conjugate and $T$ is the length of the time interval. Note that $\langle . \rangle$ represents an ensemble average. The power spectrum of the voltage time series is obtained in the same way, replacing in (\ref{Eq:spk_train_spectrum}) the spike train $x_j(t)$ by the voltage time series $v_j(t)$. In the case of the spike train power spectrum, the units are $1 / \rm s$ whereas the units of the voltage spectrum are $\rm mV^2/ \rm Hz$.

An average over a subset that includes $K$ neurons
renders the average power spectrum: 
\begin{equation}
\bar{S} = \frac{1}{K} \sum_{j \in K} S_{xx,j}(f) ,
\label{Eq:avg_spk_train_spectrum}
\end{equation}

We quantify the degree of oscillatory activity in the network via the spectral entropy $H_s$ \citep{blanco2013,kumar2016}. Spectral entropy is computed from the time-dependent firing rate (\ref{Eq:firing_rate}) as 

\begin{equation}
H_s = \frac{-\sum_k S_{rr}(f_k) \log S_{rr}(f_k)}{\log N_b},  
\label{Eq:spktral_entropy}
\end{equation}

where $N_b$ is the number of frequency bins and $S_{rr}(f_k)$ is the value of the normalized (i.e. $\sum_k S_{rr}(f_k) = 1$) power spectrum of the network time-dependent firing rate $r(t; \Delta t)$ at the $k$th bin. 
In  our simulations we use $N_b$=1000. In the case of broadband noise activity, the power spectrum of the network firing rate is flat and the spectral entropy is maximal: $H_s=1$. If, in contrast, all power is concentrated at one frequency, a case of single-frequency network oscillations, the spectral entropy vanishes: $H_s=0$.

To quantify the degree of synchrony in the network, we use the phase locking value ($PLV$) which is a standard measure to evaluate phase synchronization \citep{lachaux1999,celka2007,rosenblum2011,aydore2013,lowet2016}. Unless otherwise stated, the time average used to calculate the $PLV$ is always taken over a simulation interval of $T=$ 2000 ms.
We define the $PLV$ as the average over $K$ neuron pairs and $T$ sample time points: 
\begin{equation}
PLV = \frac{1}{K} \sum_{\{ij\}}^K\left|\sum_{t}^T e^{i \Delta \Phi_{xy}(t)} \right|,   
\label{Eq:sync_index_plv}
\end{equation} 
where $\Delta \Phi_{xy}(t)$ are the phase differences $\rho_x\Phi_{x}(t)-\rho_y\Phi_{y}(t)$ from two randomly chosen spike-trains $\left(x(t),y(t)\right)$ that are obtained using the Hilbert transform. The values $\rho_x$ and $\rho_y$ define the frequency ratio and, expecting similar firing rates, we set $\rho_x=\rho_y=1$. The $PLV$ is bounded between 0 (asynchrony) and 1 (synchrony).

In our simulations, we constructed parameter space plots of the synchrony index $PLV$
(like the ones shown in Results) for different numbers $K$ of neuron pairs and observed a saturation in the plots for increasing values of $K$ above 50. This indicates that $PLV$ becomes independent of the number of neuron pairs for $K \geq 50$. To ensure this independence, in computations we took $K=60$. 

Numerical integration of the differential equations was performed by means of the Heun algorithm \citep{mannella2002}. We used C++ to write the computational code, and Matlab and xmgrace to visualize and analyze the results. 

\section {Results}

\subsection {Preliminaries and the deterministic setup}
\label{subsec:deterministic_setup}

To single out the effects caused by the introduction of synaptic noise, we first characterize the system in the non-perturbed state, i.e. in the absence of noise. Below, we refer to this case as the deterministic setup. 

At the chosen parameter values the global state of rest is stable.
Since in the deterministic setup no activity can be excited from that state without an initial disturbance, we start simulations by applying brief electric stimulation to arbitrarily selected neurons. Different stimuli are constructed by varying
\begin{itemize}
\item the amplitude of the input current from $I_{\rm stim} = 10$ to $I_{\rm stim} = 20$;
\item the duration of the input current from $t_{\rm stim} = 50$ ms to $t_{\rm stim} = 300$ ms; and
\item the proportion of stimulated neurons: 1, 1/2, 1/4, 1/8, 1/16.
\end{itemize}

The initial kick provided by brief stimulation has a sole role to put the system into a state other than rest. After the stimulation ends, the network is left to evolve freely and its dynamics is recorded. Eventually all trials end up in the state of rest.
In most cases evolution is not a straightforward decay but a long dynamical transient; its duration strongly (by several orders of magnitude) varies, depending on the initial conditions. On discarding the cases where the free activity was shorter than $400$ ms, we are left with a set of trials in which the network displayed long-living self-sustained activity; duration of
the latter stage justifies a closer look at its intrinsic characteristics.

\begin{figure*}[!ht]
\includegraphics[scale=0.4]{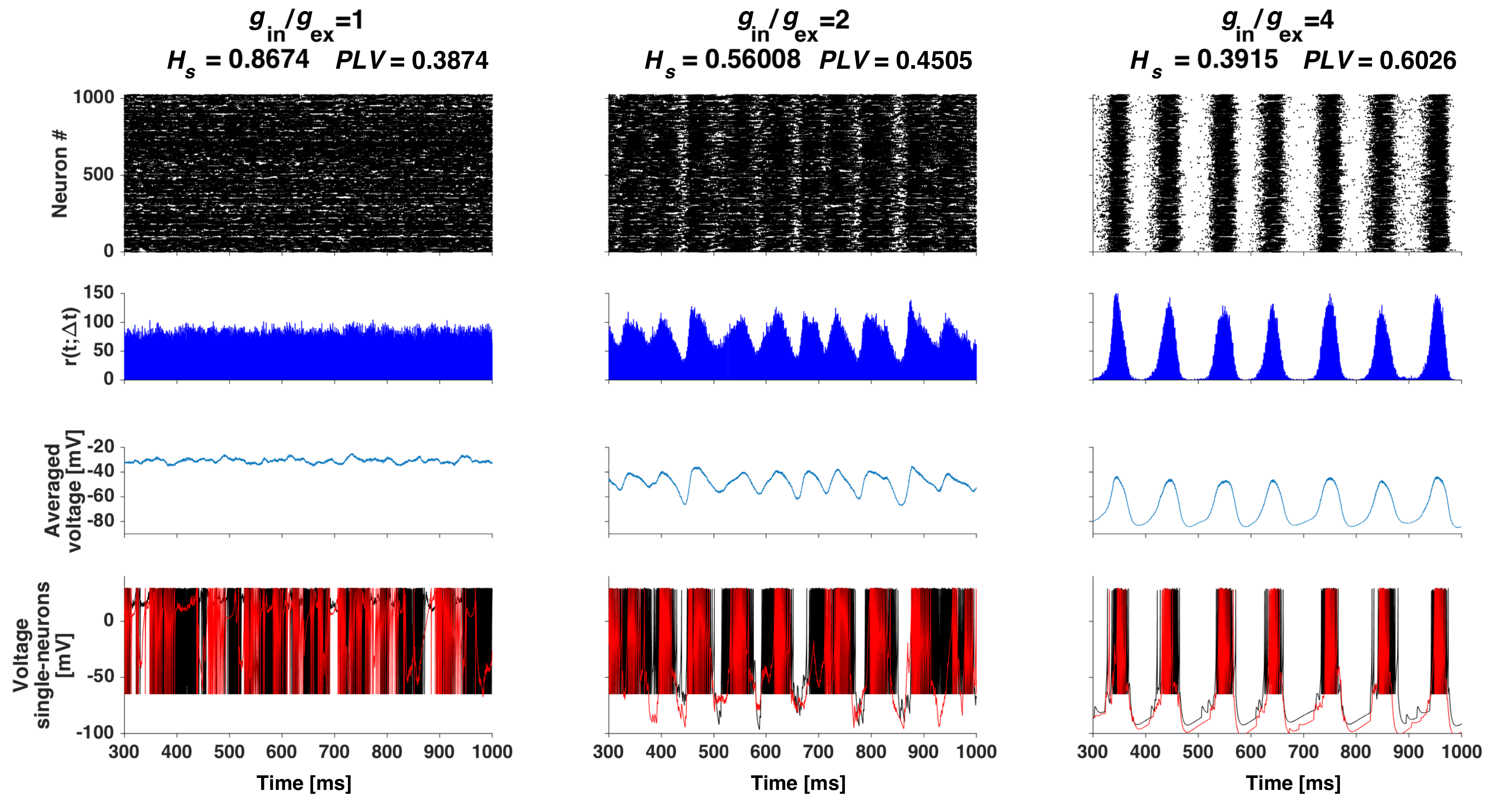}
\caption{\textbf{Self-sustained firing pattern changes under variation of $g_{\rm in}/g_{\rm ex}$ ratio in the deterministic setup}. The network is composed of RS and LTS neurons. Each column represents a combination of $g_{\rm in}/g_{\rm ex}$ indicated atop together with the corresponding spectral entropy $H_s$ and synchrony index $PLV$. From top to bottom: raster plot, network firing rate, average voltage and voltage traces of two arbitrarily selected neurons (in black and red respectively).}
\label{Fig:gex_gin_deterministic}
\end{figure*}

We have studied different combinations of the conductance increments ($g_{\rm in},g_{\rm ex}$) and observed rather distinct behavior as shown in Fig~\ref{Fig:gex_gin_deterministic}. The choice of $g_{\rm ex}$ and $g_{\rm in}$ directly affects the network balance and shapes thereby its dynamics \citep{vogels2005review}.

Depending on the ratio $g_{\rm in}/g_{\rm ex}$, the self-sustained activity displayed by the network belongs to one of two categories outlined in \cite{tomov2014}. The first one, shown in the left column of Fig~\ref{Fig:gex_gin_deterministic}, is a relatively constant network activity state where neurons spike in an asynchronous and non-oscillatory fashion. For the given example, this is confirmed by the high value of the spectral entropy ($H_s=0.87$) and the low phase locking value ($PLV=0.39$). The reason for the constant network activity can be seen from the behavior of the voltage traces for two randomly selected neurons at the bottom of the left column of Fig~\ref{Fig:gex_gin_deterministic}: the neurons fire irregularly, but their firing rates are so high that the collective activity is constant.  

The second category, shown in the right column of Fig~\ref{Fig:gex_gin_deterministic}, is an oscillatory state ($H_s=0.39$) characterized by regular periods of high mean firing rate intercalated with periods of very low firing. The average voltage indicates that the bulk of neurons is fluctuating between depolarization and hyperpolarization. The $PLV$ for this state is higher ($PLV=0.60$) indicating that neurons fire/stay silent with a higher degree of synchrony. Voltage traces for two randomly picked neurons (see bottom plot in the right column of Fig~\ref{Fig:gex_gin_deterministic}) show bursts of closely spaced spikes during high activity phases intercalated with periods of hyperpolarization below the reset value during low activity phases. This behavior was explained by us earlier \citep{tomov2016} in terms of the dynamics of the recovery variable $u$ in the single neuron phase plane of the Izhikevich neuron.

The example given in the middle column of Fig~\ref{Fig:gex_gin_deterministic} illustrates the transition between the two above categories. This transition occurs when the inhibitory synaptic increment overcomes the excitatory synaptic increment as reported in \cite{tomov2014}. The network activity in the transition region looks as a mixture of constant and oscillatory activity, with intermediate values of the synchrony ($PLV=0.45$) and oscillatory activity ($H_s=0.56$) indexes. Voltage traces for two randomly chosen neurons (bottom of middle column) show high firing rates like in the first category (a tendency for constant activity), but now there are short periods of activity break like in the second category (oscillatory activity).  

Naively, $g_{\rm in}/g_{\rm ex}=4$ may seem to be a balanced situation, as reported elsewhere \citep{brunel2000dynamics}. Here, however, we are dealing with neurons from different electrophysiological classes, and their firing rates differ as well. In addition, we are using a conductance based synaptic model where the synaptic current is voltage-dependent. In that sense,
the mean time-averaged synaptic input for a given neuron $j$ can be roughly estimated as 
\begin{eqnarray}
\label{Eq:mean_inpu}
I_j(t) \approx g_{\rm ex}C_{\rm ex}\nu_{\rm ex} \tau_{\rm ex} (E_{\rm ex} - \langle v \rangle) \\ 
\nonumber - g_{\rm in}C_{\rm in}\nu_{\rm in} \tau_{\rm in} (E_{\rm in} - \langle v \rangle),
\end{eqnarray}

\noindent where $C_{\rm ex/in}$ are the numbers of excitatory/inhibitory inputs to neuron $j$, $\nu_{\rm ex/in}$ are the mean firing rates of the excitatory/inhibitory populations, and $\langle v \rangle$ is a representative voltage. The expression in Eq.(\ref{Eq:mean_inpu}) elucidates that the notion of ``balance'' is subtle, and its reduction to just $g_{\rm ex/in}$ and $C_{\rm ex/in}$ may be misleading. Usually, when LIF neurons are considered, equal mean firing rates of excitatory and inhibitory neurons, as well as equal relaxation times $\tau_{\rm ex,in}$ are assumed, hence the balance requires only that $g_{\rm in}/g_{\rm ex}=C_{\rm ex}/C_{\rm in}$,
which, in the widely studied situation with the number of excitatory connections four times higher, results in $g_{\rm in}/g_{\rm ex}=4$. In contrast, in a network like ours, with $\nu_{\rm in}>\nu_{\rm ex}$, there is no balance at $g_{\rm in}/g_{\rm ex}=4$, instead there is a voltage dependent input current: if $\langle v \rangle$ is depolarized (hyperpolarized), negative (positive) currents drive the neuron.

Altogether, these preliminary examples confirm that the deterministic setup, depending on the ratio $g_{\rm in}/g_{\rm ex}$, is able to generate oscillatory or constant activity. In the following, we concentrate on the oscillatory situation, when inhibition overcomes excitation. 

\begin{figure*}[!ht]
\includegraphics[scale=0.35]{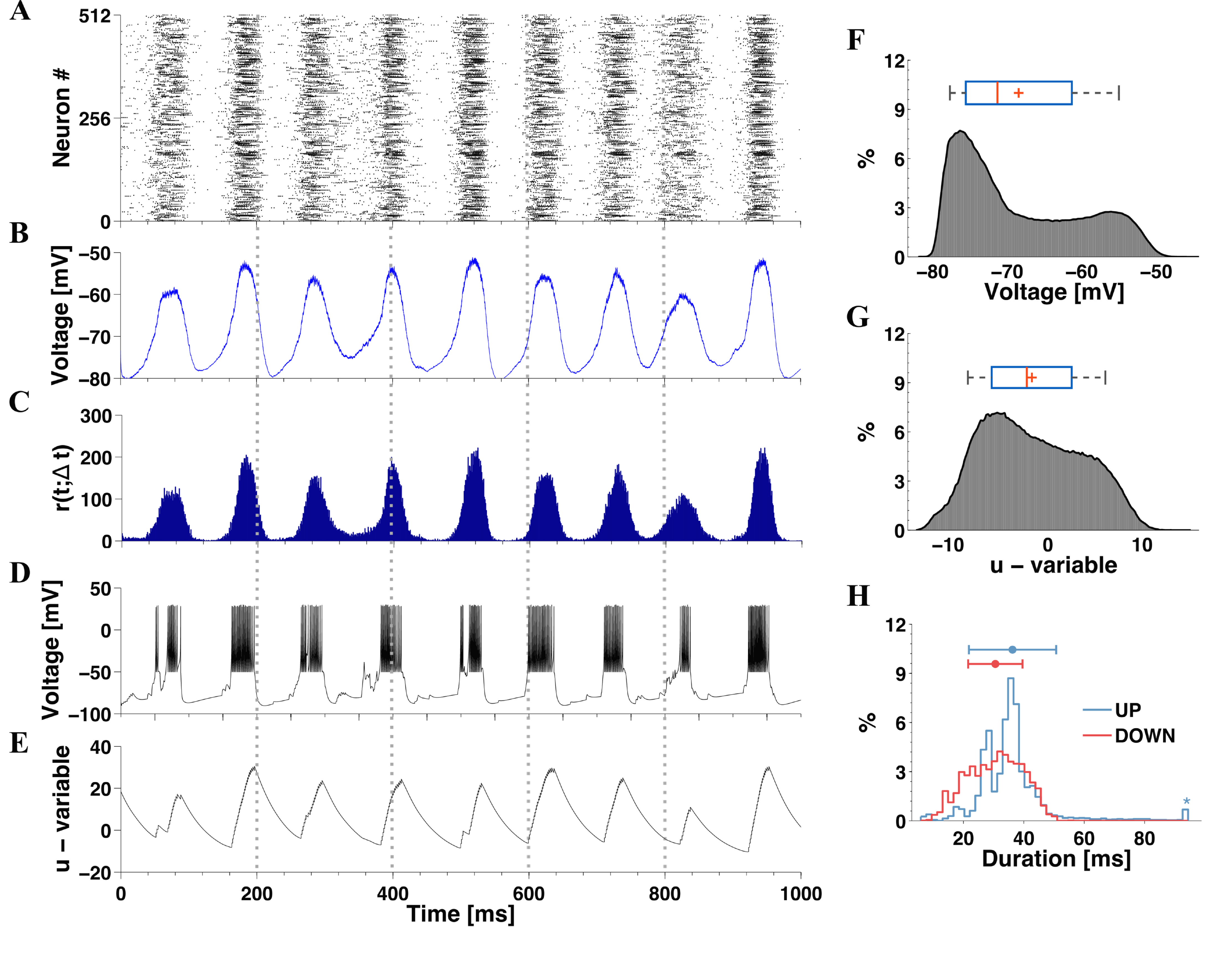}
\caption{\textbf{Up and down network oscillations in the noiseless case when $g_{\rm in} > g_{\rm ex}$.}
The network is composed of 16$\%$CH, 64\%RS and 20\%LTS neurons, with $(g_{\rm ex},g_{\rm in}) = (0.15,1)$. Panels \textbf{A-C} show the raster plot for half of the neurons in the network, average voltage and time-dependent firing rate from a sample simulation with long-lived self-sustained activity. Panels \textbf{D-E} show the voltage $v$ and membrane recovery variable $u$ extracted from a sample neuron in this simulation. Histograms \textbf{F-G}  show the distributions of average $v$ and average $u$ based on data from all long-lived simulations. In the box plots above the histograms the red lines and the pluses denote, respectively, the median and the mean. Histogram \textbf{H} presents the distribution of stay duration in the collective up and down states based on all simulations, as well as mean and standard deviation; the outlier is indicated by the star in the end of the distribution.}
\label{Fig:raster_stats_deterministic}
\end{figure*}

In Fig~\ref{Fig:raster_stats_deterministic} we present an exemplary simulation in the deterministic setup and extended statistics from the set of long-lived realizations with synaptic increments $g_{\rm ex} = 0.15$ and $g_{\rm in} = 1$ (this set contains 487 simulations, thus allowing good statistics). In this case the majority of neurons oscillates between a depolarized state and a hyperpolarized state, well visible in Fig~\ref{Fig:raster_stats_deterministic} \textbf{B} and on the bimodal distribution in Fig~\ref{Fig:raster_stats_deterministic}\textbf{F}, computed from the entire set of simulations with varied initial stimulation. For individual neurons these preferred subthreshold membrane potentials are known
as ``up'' and ``down'' states \citep{wilson2008}, and in the context of the ensemble of neurons it seems natural to view these two states as collective ``up'' and ``down'', respectively. 
As seen in Fig~\ref{Fig:raster_stats_deterministic} \textbf{A-E}, a typical period of oscillations is close to 100 ms 
($f \approx 10$ Hz).

For a typical neuron in the ensemble, Figs~\ref{Fig:raster_stats_deterministic} \textbf{D-E} illustrate the temporal evolution of the voltage and the membrane recovery variable, respectively, during the same simulation. There is strong correlation between firing of this neuron and the periods of high activity of the whole network, although some other neurons also fire when the network activity is low. These latter are inhibited during the high activity epochs and become disinhibited when the overall network activity is low. We have shown elsewhere the importance of this disinhibitory effect to sustain the long-lived activity of the network in the oscillatory situation \citep{tomov2016}. 

Remarkably, not only the voltage series in Fig~\ref{Fig:raster_stats_deterministic} \textbf{D} features two different states (a hyperpolarized one and a depolarized one) but also the membrane recovery variable, which clearly grows when the network activity is high and slowly relaxes when the activity is low. This is a global phenomenon: in all simulations there are peaks of the variable $u$.
In the distribution shown in Fig~\ref{Fig:raster_stats_deterministic} \textbf{G}, the maximum is broad due to the time-scale separation of the variables: $u$ is slower than $v$ and its relaxation takes much longer. In \cite{tomov2016} we have shown the importance of the recovery variable for the breakdown of global high-activity epochs, which produces the up and down oscillatory pattern. 

Fig~\ref{Fig:raster_stats_deterministic} \textbf{H} presents a histogram of durations in collective up and down states. The term ``up'' refers here to different states in which the network activity is above 20\% of its average value, whilst the voltage for the majority of neurons is at a depolarized value. A collective ``down'' state is identified whenever the bulk of the neurons reaches a hyperpolarized state close to $-80$ mV.

Recall that eventually the system ceases to oscillate, and voltages of all neurons invariably converge to the rest value.

\subsection {Setup with synaptic noise}
Introduction of synaptic noise drastically changes one important aspect, both in the individual and in the collective dynamics: the state of rest, albeit formally stable, ceases to be the ultimate attractor. A neuron is an excitable system, and in the noisy setup it is just a matter of time when a sufficiently strong fluctuation (or a cumulative effect of many fluctuations) drives it across the spiking threshold. For an ensemble this implies disordered sporadic firing of its members, which, under favorable conditions, can turn into ordered collective activity. If deterministic aspects dominate in dynamics, this activity will temporarily end in the state of rest, only to be recreated by new fluctuations.
\subsubsection{Isolated neurons}
\label{sec::isolated}

Consider an individual neuron that obeys Eq.(\ref{eq:Izh-neuron}) with the synaptic current $I$ given by Eq.(\ref{Eq:current})
and synaptic conductances $G^{\rm ex/in}$ governed by Eq.(\ref{Eq:synapse}) with noisy input. An isolated neuron, by definition, has no synaptic inputs; nevertheless, stochastic fluctuations of its synaptic conductances can result in action potentials.
In this situation, to study the influence of noise on the resting neuron we, without loss of generality, 
set $n_j=1$ in Eq.(\ref{Eq:synapse}).
Take the initial conditions for the neuron at its state of rest and set its synaptic conductances to zero, so that the initial current is absent. 
As time goes on, the conductance evolves stochastically; to ensure that it stays positive, we impose a reflecting condition at zero (which, in the long run, very
slightly shifts upwards the mean value of $\xi(t)$).
As a result, a stochastic current $I(t)$ is generated. As long as $I(t)$ is absent or 
sufficiently small, the neuron stays at rest. As soon as the instantaneous current $I$ exceeds the critical value 
$I_{\rm crit}(t)=\displaystyle\frac{(\beta-b)^2}{4\alpha}-\gamma$, 
with $\alpha,\beta,\gamma,b$ being the parameters of the Izhikevich model (\ref{eq:Izh-neuron}), the state of rest disappears (the mechanism is explained below in Sect. \ref{subsection_single}),
the voltage variable $v$ starts to grow monotonically, and the neuron fires.

Since presynaptic inputs are absent in this isolated neuron description (see Eq.~ \ref{Eq:synapse}), computation of the first firing time for an isolated neuron turns into a variant of the mean first passage time problem \citep{siegert1951} for the Ornstein-Uhlenbeck process. Numerically, we find this quantity by averaging over a sufficient number of trials.
 
\begin{figure}[!ht]
\centerline{\includegraphics[scale=0.3]{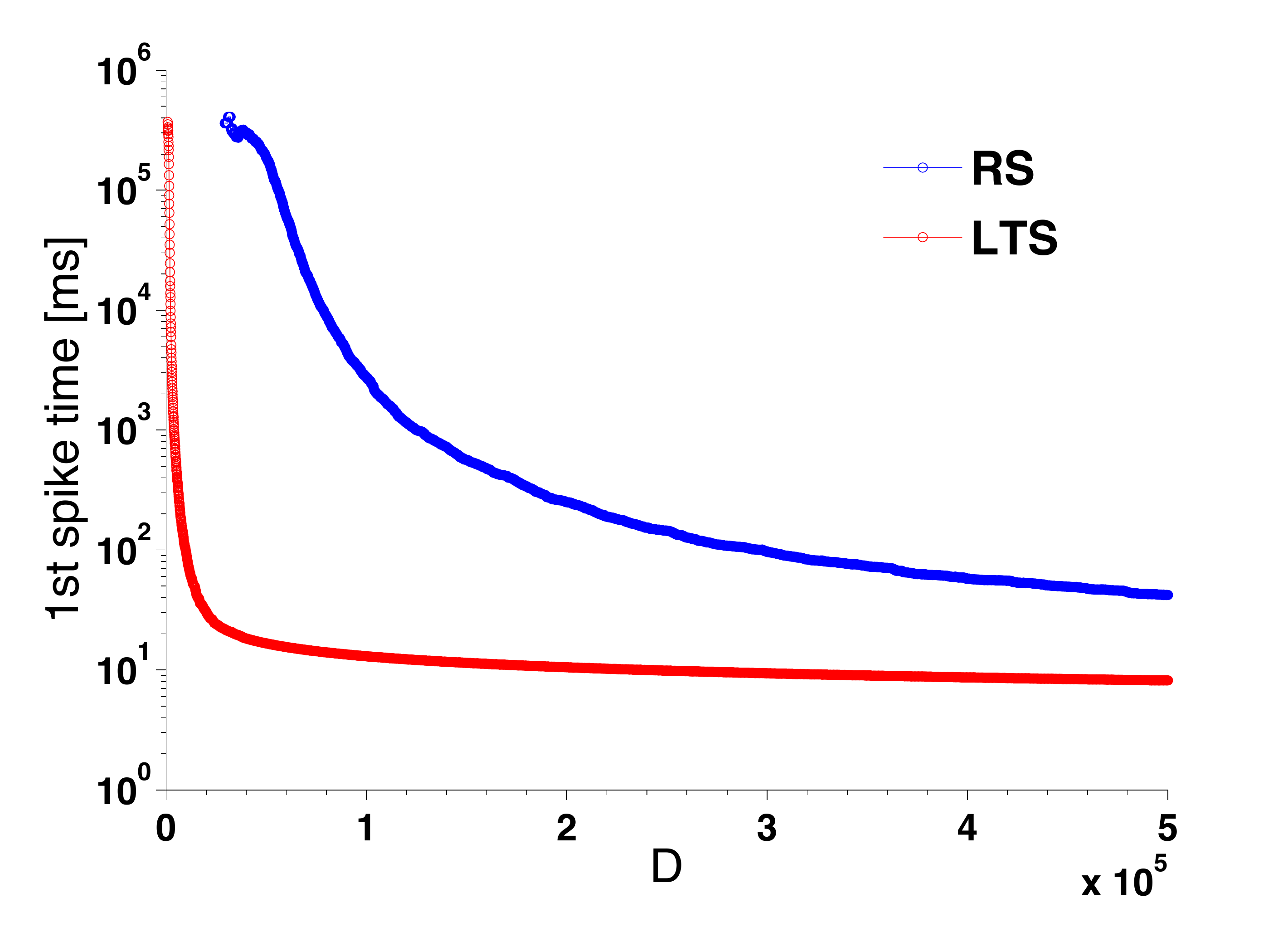}}
\protect \caption{\textbf{Average time of first spike for the Izhikevich neuron model driven by synaptic noise.} $D$: noise intensity. Blue curve: RS neuron. Red curve: LTS neuron.}
\label{Fig:singleneuron_noise}
\end{figure}

Regarding dependency of $I_{\rm crit}(t)$ on the electrophysiological class, we note that the parameters $\alpha$, $\beta$ and $\gamma$ are common for all classes, leaving $b$ as the only parameter that matters. In this context, $b$ determines the current threshold value. Furthermore, three of the four considered neuronal classes share the same value of $b$, whereas the LTS neuron has a higher value of $b$, ensuring early initiation of spikes. Hence, 
it suffices to compare two neurons: LTS and e.g. RS. In Fig~\ref{Fig:singleneuron_noise} we plot the computed dependences of the time of first spike on the synaptic noise intensity.

Notably, from the point of view of the random network, each curve in Fig~\ref{Fig:singleneuron_noise} shows the behavior for all neurons of its respective kind, regardless of their in-degree: according to Eq.(\ref{Eq:synapse}), an increase of the in-degree (in other words, of the number of independent Gaussian noises acting upon the synapse) rescales the variance and is therefore equivalent to the corresponding increase of $D$ at constant degree. Recall that in the studied networks most of the neurons have in-degree $\approx 10$. Altogether, the influence of the number of synaptic connections is clear: the higher the in-degree, the higher the variance of the input noise, the faster the neuron crosses the threshold and emits a spike. 

\subsubsection{Network with weak synaptic noise}
\label{subsec:weak_noise}

\begin{figure*}[!ht]
\includegraphics[scale=0.4]{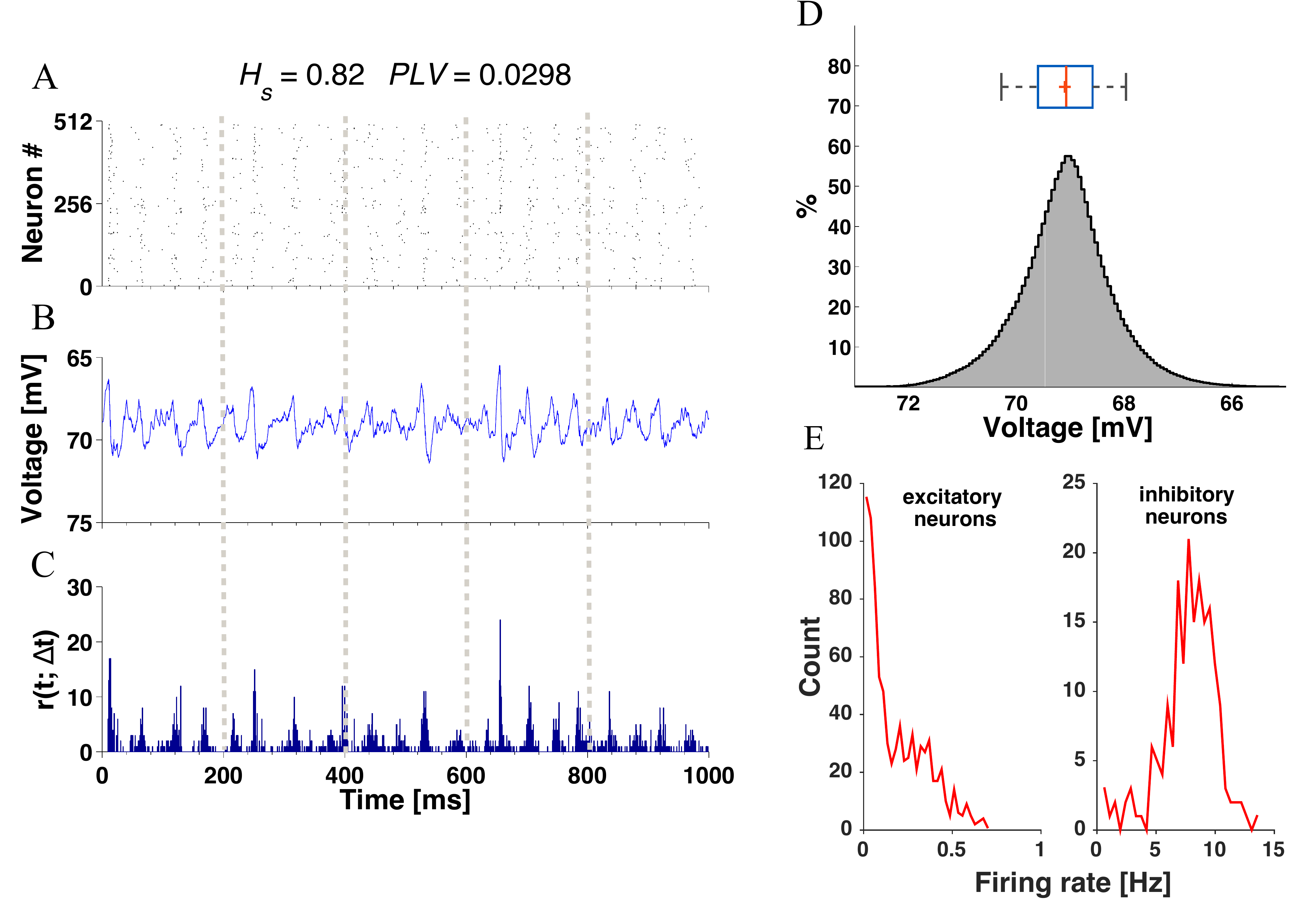}
\caption{\textbf{Asynchronous irregular state in the presence of weak synaptic noise.}
The network, composed of 16$\%$CH, 64\%RS and 20\%LTS neurons, evolves without initial stimulation. Synaptic increments: $(g_{\rm ex},g_{\rm in}) = (0.15,1)$. Intensity of synaptic noise: $D=2.5\times10^{-6}$. Panels \textbf{A-C} present, respectively, raster plot for half of the neurons in the network, average voltage and time-dependent firing rate for the network. Above them  the values of $H_s$ and $PLV$ are cited. Panels \textbf{D-E} are histograms with distributions of average voltage and firing rates. For the firing rates, excitatory and inhibitory populations are presented separately, as indicated in the titles of \textbf{E}.}
\label{Fig:low_noise}
\end{figure*}

We begin the discussion of synaptic noise in the network by presenting a case where its introduction induces activity 
with properties strongly different from those in the deterministic setup. For the same set of parameter values as in the deterministic case of Fig~\ref{Fig:raster_stats_deterministic}, instead of initial stimuli, we add in accordance with  Eq.(\ref{Eq:synapse}) small ($D=2.5\times10^{-6}$) stochastic fluctuations to the synaptic variables. 
This results in activity with very low firing rates, exemplified in panels \textbf{A-C}
of Fig~\ref{Fig:low_noise}. The high spectral entropy ($H_s = 0.82$) and the very low synchrony ($PLV = 0.0298$) indicate a non-oscillatory and asynchronous type of activity. The voltage distribution in Fig~\ref{Fig:low_noise} \textbf{D} stands in contrast to the deterministic case: it is unimodal, the maximum lies at the mean, and the relevant voltage values are close to the resting potential. The firing rates in Fig~\ref{Fig:low_noise} \textbf{E} are close either to $1$ Hz (excitatory neurons) or to 8 Hz (inhibitory neurons). The state of the network in the weak synaptic noise setup corresponds well to the so-called asynchronous irregular (AI) state \citep{brunel2000dynamics,vogels2005review}. 

Up-down oscillations can occur in the weak synaptic noise setup, but only for short transient periods like in the deterministic case. After the transient, the persistent activity is asynchronous irregular like the one in Fig~\ref{Fig:low_noise}. An example is shown in \nameref{S1_Fig}.

\subsection {Onset and classification of intermittent oscillatory and quiescent activity in the synaptic noise setup}

\begin{figure*}[!ht]
\includegraphics[scale=0.48]{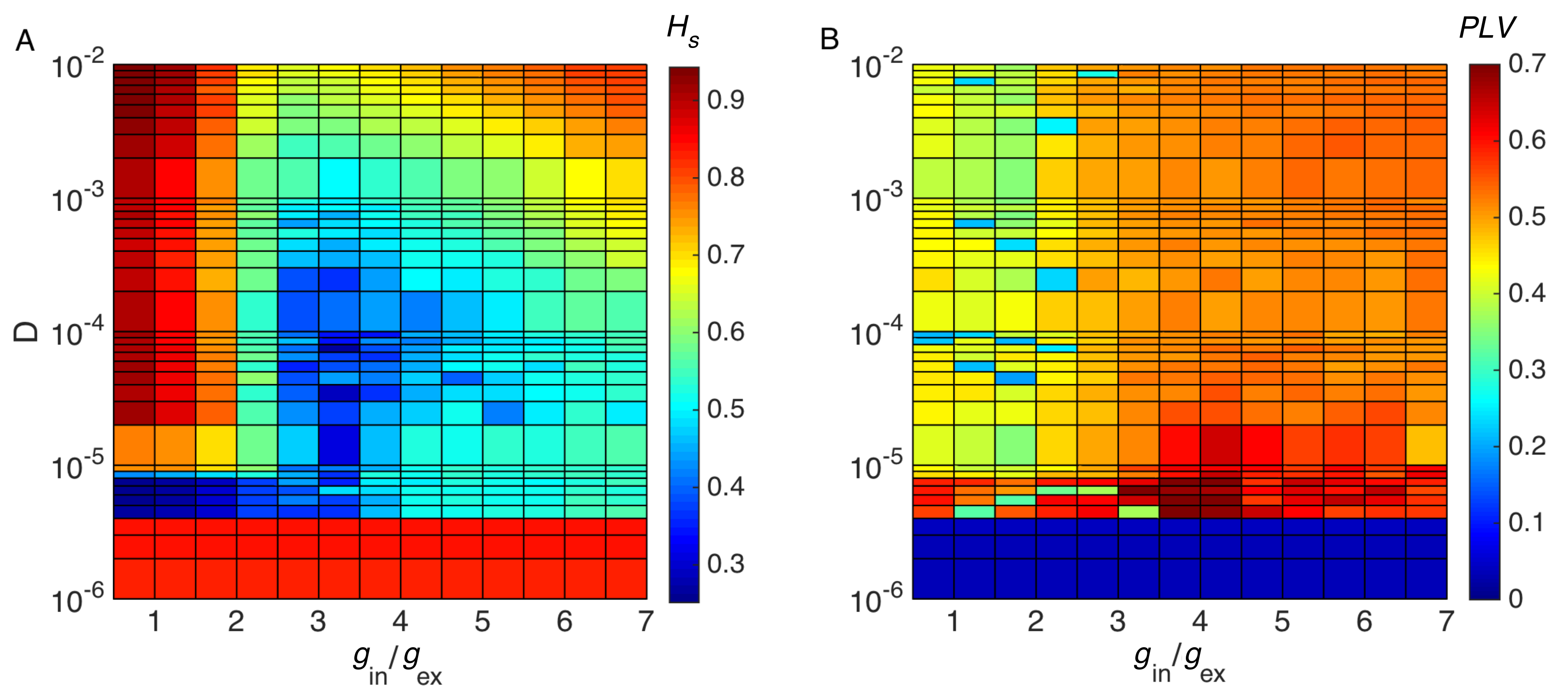}
\caption{\textbf{Spectral entropy $H_s$ and synchrony index $PLV$ for the synaptic noise setup.} Two-dimensional space where ordinate represents the synaptic noise intensity $D$ and abscissa, the ratio of synaptic increments $g_{\rm in}/g_{\rm ex}$. The coordinate mesh is linear (from 0.5 to 7) with respect to $g_{\rm in}/g_{\rm ex}$ and logarithmic with respect to the synaptic noise intensity (from $D=1\times10^{-6}$ to $D=1\times10^{-2}$). Panel \textbf{A}: Colors represent spectral entropy $H_s$ (values close to zero correspond to oscillatory states and values close to 1 correspond to non-oscillatory states). Panel \textbf{B}: Colors represent synchrony evaluated by means of the phase locking value $PLV$ (values close to zero correspond to asynchronous states whereas values close to 1 correspond to synchronous state).}
\label{Fig:diagram_rates}
\end{figure*}

Here we describe various collective states induced in the network by synaptic noise. Experience gained from the study of the deterministic setup allows us to expect that, along with the synaptic noise intensity $D$, the crucial parameter in this context is the ratio $g_{\rm in}/g_{\rm ex}$: the proportion between inhibitory and excitatory synaptic strengths \citep{brunel2000dynamics,girones2015}. We start by exploring the behavior of the spectral entropy $H_s$ and the synchrony measure $PLV$ in the two-dimensional diagram spanned by parameters $g_{\rm in}/g_{\rm ex}$ and $D$ (Fig~\ref{Fig:diagram_rates}).  

As seen in the diagrams in Fig~\ref{Fig:diagram_rates}, both $g_{\rm in}/g_{\rm ex}$ and $D$ are responsible for shaping the activity pattern of the network. Let us begin with the diagram for spectral entropy in Fig~\ref{Fig:diagram_rates} \textbf{A}. For weak synaptic noise ($D \lessapprox 5 \times 10^{-6}$) the network displays non-oscillatory behavior independently of the  ratio $g_{\rm in}/g_{\rm ex}$. For the narrow horizontal band defined by $5 \times 10^{-6} \lessapprox D \lessapprox 10^{-5}$, the state of the network is oscillatory and the degree of oscillatory activity is higher for $g_{\rm in}/g_{\rm ex} \lessapprox 2$. On the other hand, for $D \gtrapprox 10^{-5}$ the situation is inverted and the region determined by $g_{\rm in}/g_{\rm ex} \lessapprox 2$ displays non-oscillatory activity while most of the remainder of the diagram features oscillatory activity. Within this latter part of the diagram, 
increase of both noise and inhibitory synaptic strength lowers the degree of oscillatory activity  until in the upper right corner the activity turns non-oscillatory.

Now let us turn to the diagram for the synchrony $PLV$ in Fig~\ref{Fig:diagram_rates} \textbf{B}. The region of
weak synaptic noise  ($D \lessapprox 5 \times 10^{-6}$) displays asynchronous behavior independently of $g_{\rm in}/g_{\rm ex}$. Under such weak noise firing remains an individual event for noise-perturbed neurons, rather than a collective effect. Along the narrow horizontal band of the diagram determined by $5 \times 10^{-6} \lessapprox D \lessapprox 10^{-5}$, the synchrony index has mostly intermediate values with a narrow high-synchrony region around $g_{\rm in}/g_{\rm ex} \approx 4$. In the remainder of the diagram the behavior along horizontal scans in the diagram is roughly the same: in the entire region determined by $g_{\rm in}/g_{\rm ex} \lessapprox 2.5$ the activity is asynchronous, whereas outside that region the degree of synchrony has intermediate values. 

\begin{figure*}[!ht]
\includegraphics[scale=0.33]{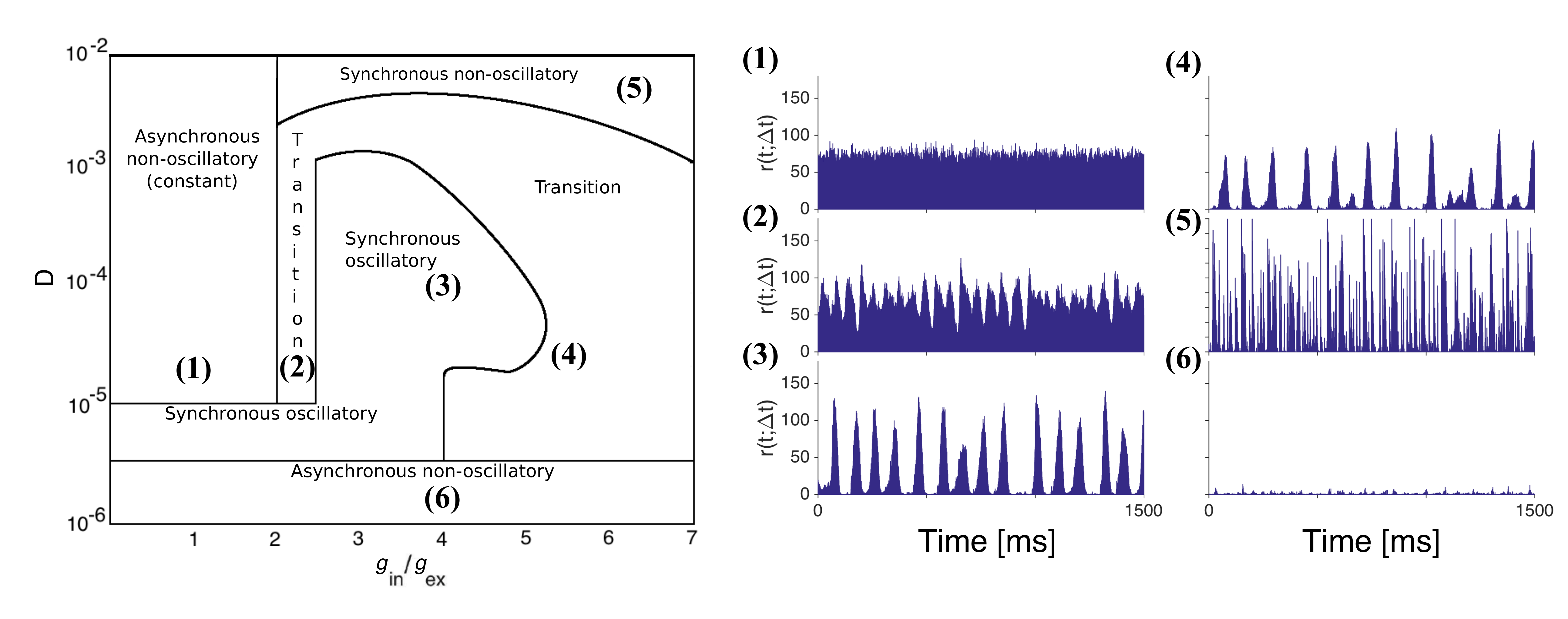}
\caption{\textbf{Network activity patterns in the synaptic noise setup.} A schematic representation of the $D$ vs. $g_{\rm in}/g_{\rm ex}$ diagram of Fig~\ref{Fig:diagram_rates} combining the information on degree of oscillatory activity ($H_s$) and degree of synchrony ($PLV$) disclosed in that figure. The names of the activity types are given inside the regions bounded by full lines. The synchronous non-oscillatory type is equivalent to the constant type used to describe network states in the deterministic setup. The region marked as ``transition" corresponds to states with intermediate levels of oscillatory activity and synchrony. On the right side of the diagram we present the time-dependent firing rate $r(t;\Delta t)$ of the network for six selected ($D,g_{\rm in}/g_{\rm ex}$) combinations. Numbers on the left-hand top of the panels indicate the corresponding points in the diagram to the left.}
\label{Fig:combined}
\end{figure*}

The combined information in the two diagrams of Fig~\ref{Fig:diagram_rates} is qualitatively summarized in a schematic diagram drawn in Fig~\ref{Fig:combined}. The 
states in this diagram are denoted in accordance with
two measures of network activity in Fig~\ref{Fig:diagram_rates}: $H_s$ quantifies the degree of oscillatory activity and $PLV$ quantifies the degree of synchrony. Selected samples from the different regions are also displayed on the right of Fig~\ref{Fig:combined} to show the time-dependent network firing rates for the corresponding combinations of $H_s$ and $PLV$.

The region of weak synaptic noise lies at the bottom of the diagram in Fig~\ref{Fig:combined}. The type of network activity there is asynchronous non-oscillatory, already described in subsection~\ref{subsec:weak_noise}. It is similar to the asynchronous irregular (AI) activity observed in networks of LIF neurons \citep{brunel2000dynamics,vogels2005review}. The region stretches along the full length of the horizontal axis, indicating that the generic features of the network activity for weak synaptic noise are insensitive to the ratio between excitation and inhibition.   

For stronger synaptic noise the structure of the diagram in Fig~\ref{Fig:combined} is more complex. 
The network displays synchronous oscillatory activity within an irregular shaped region in the center of the diagram, adjoined by a narrow horizontal strip in the bottom part. This is similar to the synchronous regular (SR) type of activity found in networks of LIF neurons \citep{brunel2000dynamics,vogels2005review}. In the remainder of the third of the diagram where $g_{\rm in}/g_{\rm ex} < 2$ the activity is asynchronous non-oscillatory. Its pattern is similar to the constant pattern shown in the left column of Fig~\ref{Fig:gex_gin_deterministic}. On the other hand, through the upper two-thirds of the diagram for $g_{\rm in}/g_{\rm ex} > 2$ the activity is synchronous non-oscillatory. Thus, for very strong synaptic noise the network activity is non-oscillatory and can be synchronous or asynchronous depending on the $g_{\rm in}/g_{\rm ex}$ ratio. 

Finally, the diagram in Fig~\ref{Fig:combined} includes the region marked as ``transition". It contains most of the right third of the diagram, with the exception of the regions of weak and strong synaptic noise mentioned above, and extends to the central part of the diagram 
where it separates the synchronous oscillatory from the asynchronous non-oscillatory regions. This corresponds to a region with intermediate degrees of oscillatory activity (the greenish region in the diagram for $H_s$ in Fig~\ref{Fig:diagram_rates} \textbf{A}) and synchrony (red-orange to yellow-orange colors in the diagram for \emph{PLV} in Fig~\ref{Fig:diagram_rates} \textbf{B}). Therefore, states in the transition region should occupy intermediate position between constant and oscillatory states like the state in the middle column of Fig~\ref{Fig:gex_gin_deterministic}. 

Interested in the behavior of the network in the transition region, we focus here on a part of the diagram in Fig~\ref{Fig:combined} determined by $(g_{\rm in},g_{\rm ex})=(1,0.15)$, which implies $g_{\rm in}/g_{\rm ex}\approx 6.66$, and $10^{-5} \lessapprox D \lessapprox 10^{-4}$. This corresponds to the greenish (light orange) region on the lower right-hand side of the diagram for $H_s$ ($PLV$) in Fig~\ref{Fig:diagram_rates} \textbf{A} (\textbf{B}). Spectral entropy and \emph{PLV} here are both close to 0.5 meaning that states with intermediate levels of oscillatory activity and synchrony may be encountered. 

\begin{figure*}[!ht]
\includegraphics[scale=0.35]{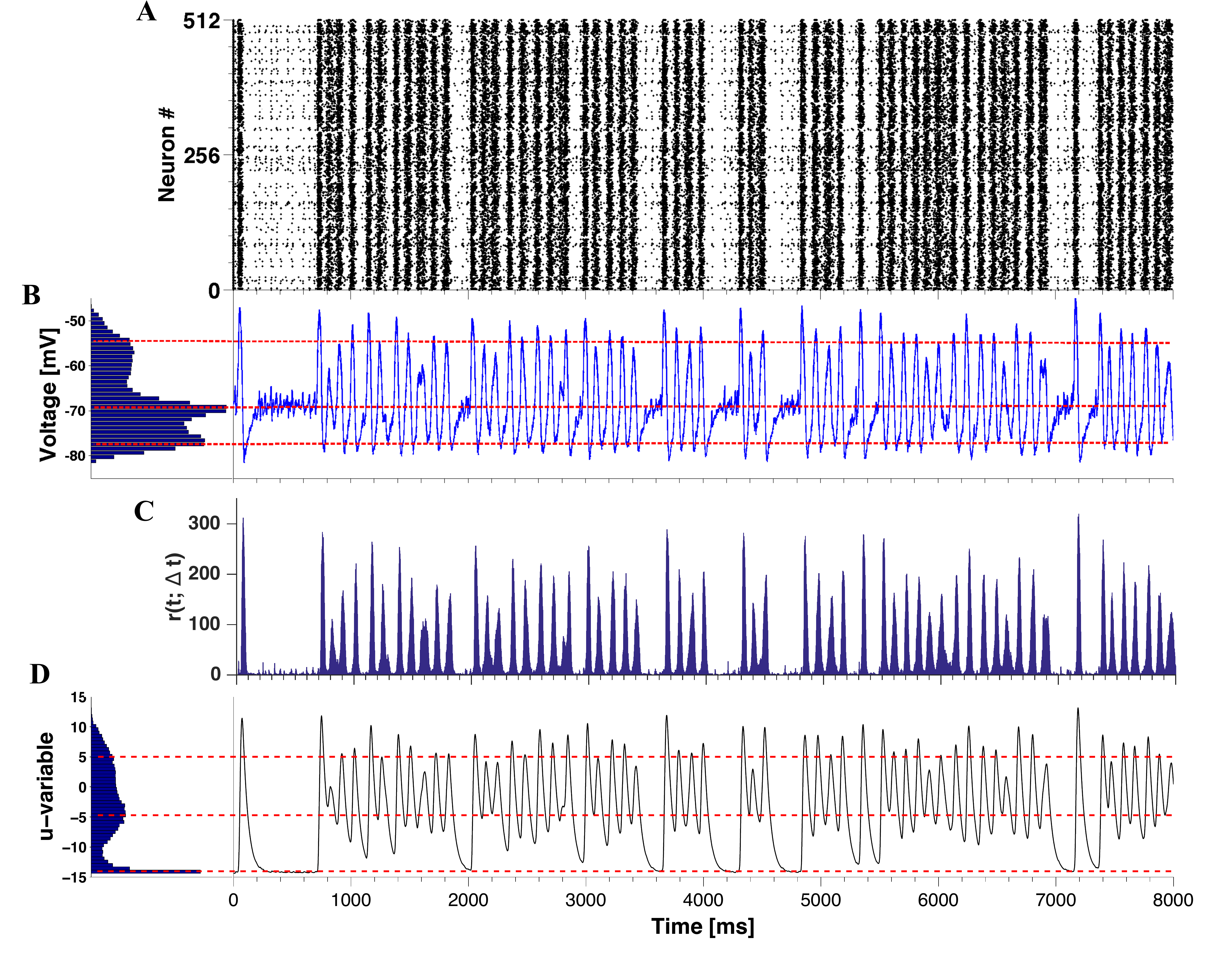}
\caption{\textbf{Intermittent transitions between active oscillatory and quiescent regimes in the presence of synaptic noise.} Plots generated for a network with 16$\%$CH, 64\%RS and 20\%LTS neurons, $D=1\times10^{-5}$ and $(g_{\rm in},g_{\rm ex})=(1,0.15)$. 
Panel \textbf{A}: Raster plot for half of the neurons in the network.  
Panel \textbf{B}: voltage $v$ histogram (left) and time course of average voltage over all network neurons (right).
Panel \textbf{C}: time-dependent firing rate of the network.
Panel \textbf{D}: Recovery variable $u$, histogram (left) and time course of average recovery value over all network neurons.}
\label{Fig:raster_sample}
\end{figure*}

In Fig~\ref{Fig:raster_sample} we illustrate dynamics for the point given by $D=1\times10^{-5}$ and $(g_{\rm in},g_{\rm ex})=(1,0.15)$ in the diagram in Fig~\ref{Fig:combined}. This point is in the transition region on the lower right-hand side of the diagram described above, which is characterized by intermediate values of $H_{s}$ and \emph{PLV}.  

Remarkably, a typical record of a long simulation trial in this region of the diagram consists of alternating states (Fig~\ref{Fig:raster_sample}): an oscillatory one, akin to oscillations presented in the deterministic setup in Fig~\ref{Fig:raster_stats_deterministic}, and a state with very low firing rates similar to the one in Fig~\ref{Fig:low_noise}. From time to time transitions between these states occur, seemingly without any precursors. Compared to deterministic simulations, an additional feature is distinct in the histogram of mean voltage: a pronounced maximum at the state of rest. Accordingly, the temporal evolution of voltage is organized around three characteristic values, instead of two known from the deterministic setup. Three red dashed lines in Figs~\ref{Fig:raster_sample} \textbf{B-C} mark three relevant states; from top to bottom, they denote depolarization, the state of rest and hyperpolarization. Note that the histogram in \textbf{B} can be viewed as a combination of the voltage histograms from Figs~\ref{Fig:raster_stats_deterministic} and \ref{Fig:low_noise}.

The average spectral entropy calculated over the quiescent/oscillatory states in Fig~\ref{Fig:raster_sample} is $H_s = 0.74$/$H_s = 0.37$, indicating non-oscillatory activity in the first case and oscillatory activity in the second one.

We classify the observed states based on two attributes: network activity and average voltage. Like previously, the average voltage series was used to detect the up and down states (see Fig \ref{Fig:raster_stats_deterministic} \textbf{H}). The states close to rest are identified through very low  network activity,

In terms of activity, we introduce the following distinction:

\begin{itemize}
\item  \textbf{quiescent period} is the time interval when the time-dependent firing rate of the network $r(t,\Delta t)$ is below its maximum by at least 20\%, and most of the single neurons have voltage values close to the resting state. During a quiescent period there can be sporadic noise-induced spikes but no collective dynamics. The state is similar to an asynchronous irregular (AI) state of networks of LIF neurons \citep{brunel2000dynamics,vogels2005} with low firing rate, and to a desynchronized cortical state as described in the Introduction.
\item \textbf{active period} is the time interval when the network exhibits oscillatory activity, alternating between high depolarized and hyperpolarized mean voltage values: collective up and down states. Such behavior can be related to the self-sustained activity developed in \emph{in vivo} cortical slice preparations and during slow-wave sleep and anesthesia \citep{steriade2001,tomov2016,sanchez-vives2017}.
\end{itemize}

These definitions, in combination with the values of the average voltage, facilitate identification of different collective states. Certain states that look very similar on the raster plot turn out to differ in typical voltage values. For instance, both the down state and the quiescent period feature in the raster plot almost no activity, but can be easily discerned in terms of the average voltage. 

\begin{figure*}[!ht]
\includegraphics[scale=0.4]{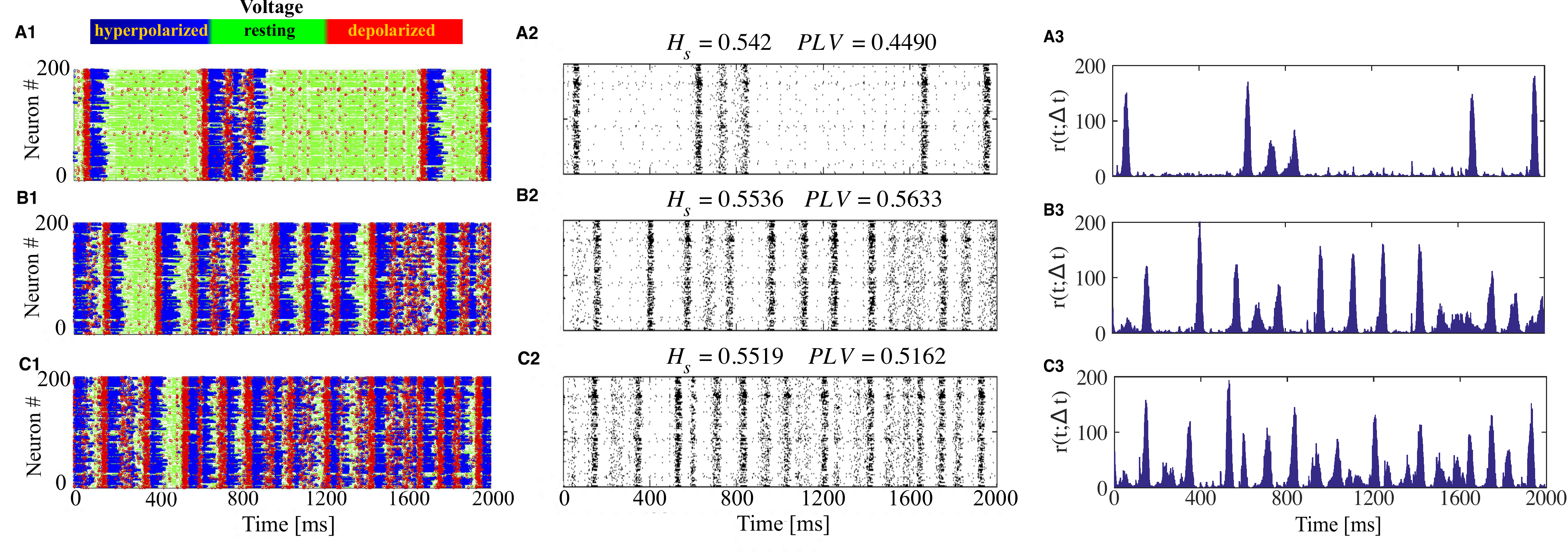}
\caption{\textbf{Increase of synaptic noise favors up-down oscillations.} The network has the same composition as in Fig~\ref{Fig:raster_sample} with varying synaptic noise intensity $D$. \textbf{A}:~$D=0.5\times 10^{-5}$, \textbf{B}:~$D=1.5\times 10^{-5}$, \textbf{C}:~$D=4.5\times 10^{-5}$. In \textbf{A1}, \textbf{B1}, and \textbf{C1} blue dots correspond to depolarization (up state), red dots to hyperpolarization (down state), and green dots to voltage near the resting state. \textbf{A2}, \textbf{B2}, and \textbf{C2}: raster plots for 200 neurons in the network with corresponding $H_s$ and $PLV$ values atop each plot. \textbf{A3}, \textbf{B3}, and \textbf{C3}: time-dependent firing rates.}
\label{Fig:raster_spktras}
\end{figure*}

In Fig~\ref{Fig:raster_spktras} we show various regimes at different values of $D$. Three samples corresponding to the time interval of $2$ s are,  from top to bottom:  $D=0.5\times 10^{-5}$, $D=1.5\times 10^{-5}$, and $D=4.5\times 10^{-5}$, respectively. In panels \textbf{A1}, \textbf{B1}, and \textbf{C1} green dots denote states with instantaneous voltage values close to the resting state, blue dots denote hyperpolarized voltage (down state), and red dots denote depolarized voltage (up state). The plot highlights the crucial role of synaptic noise level in changes of typical duration at each of these states. It is easier to generate oscillatory states (alternating between up and down states) when the network is subjected to stronger synaptic noise. In contrast, the ``green'' states close to rest (quiescent periods), prevalent at low synaptic noise amplitudes, occupy a much smaller proportion of time when synaptic noise becomes sufficiently intensive. 

Comparison of raster plots in Fig~\ref{Fig:raster_spktras} indicates that when noise intensity $D$ is increased, the waves of activity start to merge. This hinders identification of states, based on the raster plot alone. The spectral entropy and the synchrony index increase with the noise intensity. We expect that at very high levels of noise the activity becomes constant (synchronous non-oscillatory), with rather high firing frequencies (see the schematic diagram in Fig~\ref{Fig:combined}). 

\begin{figure}[!ht]
\includegraphics[scale=0.36]{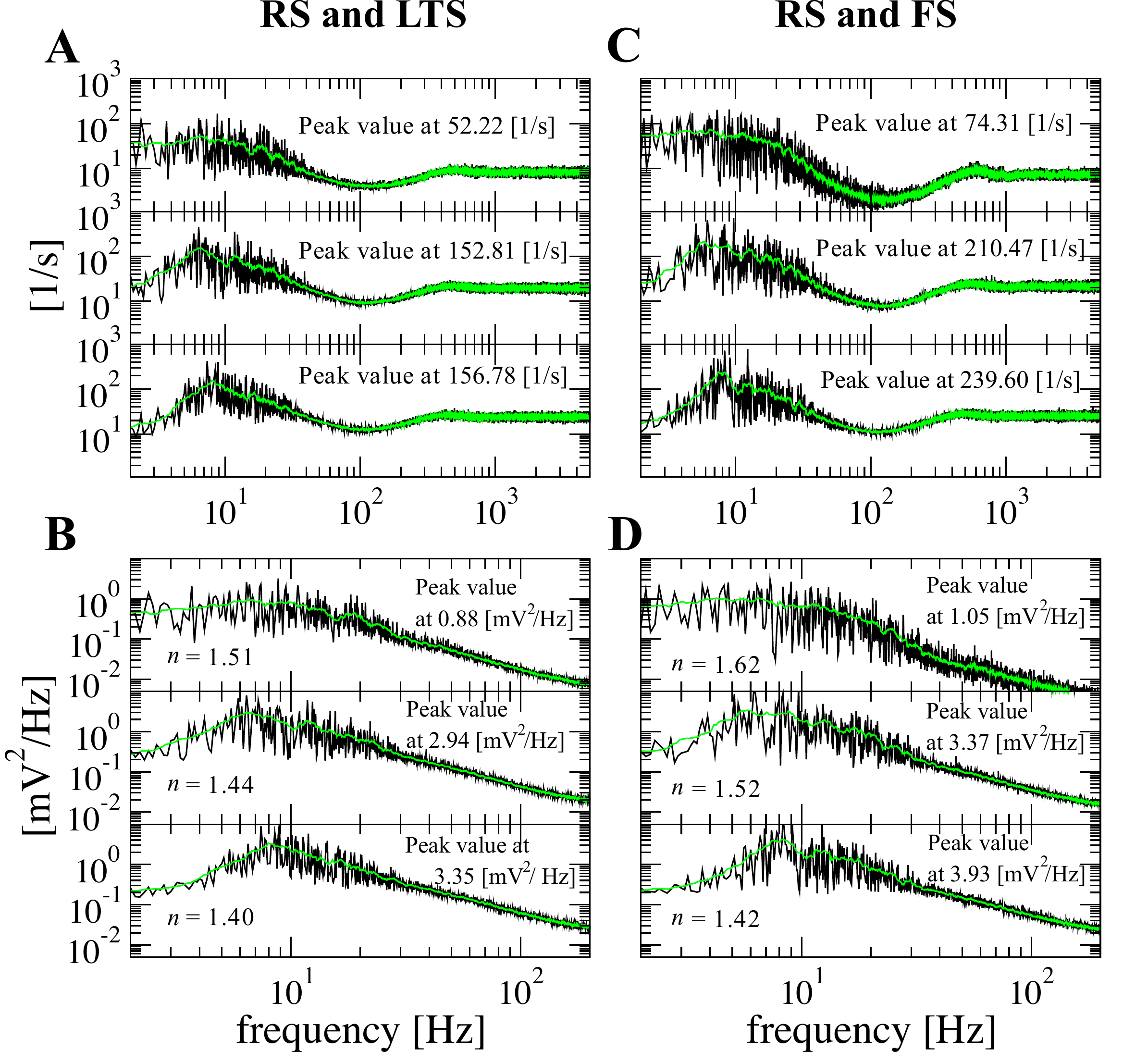}
\caption{\textbf{Averaged power spectra at different noise intensities.} 
Simulation length: 10 s.
Left column: network with inhibitory LTS neurons. Right column: network with inhibitory FS neurons. Every panel (\textbf{A},\textbf{B},\textbf{C},\textbf{D}) contains three subpanels displaying three levels of noise intensity from top to bottom: $D=0.5\times 10^{-5}$, $D=1.5\times 10^{-5}$, and $D=4.5\times 10^{-5}$.
Black curves in \textbf{A} and \textbf{C}: averaged spectra of spike trains for 200 randomly chosen neurons. 
Black curves in \textbf{B} and \textbf{D}: 
averaged spectra of voltage for the same 200 neurons. Green curves: moving average over 20 points. Peak values are indicated in the plot and were evaluated from the green curves neglecting the zeroth frequency bin. In the bottom panels we display the factor $n$ for the $1/f^n$ decay extracted between $10$ and $200$ Hz.}
\label{Fig:psd}
\end{figure}

In the frequency domain, variation of the noise level leads to redistribution of power in the Fourier spectra of both the spike trains and the voltage series.  
Fig~\ref{Fig:psd} presents spectra for the same noise intensities as in Fig~\ref{Fig:raster_spktras}: from top to bottom,
$D=0.5\times 10^{-5}$, $D=1.5\times 10^{-5}$, and $D=4.5\times 10^{-5}$. All spectra were averaged over ensembles of 200 neurons, see Eq. \ref{Eq:avg_spk_train_spectrum}. 
The shapes of spectral curves for spike trains and for voltage values are similar; the only noticeable difference is the somewhat faster decay at high frequencies in the voltage spectra. The left column shows mixtures of RS and LTS neurons; the right column corresponds to networks with RS and FS neurons. Under low levels of noise, spectral power is concentrated at very low frequencies, waves of collective activity are quite rare and, when they occur, they are mostly isolated events. On increasing the intensity $D$, waves of collective activity become more frequent whereas the periods of quiescence get shorter. During the periods of oscillatory activity, neurons are either firing at high frequency in the up state or rarely firing in the down state. This results in the increase of spectral power at low frequencies, with a distinct maximum near $10$ Hz.

Comparison of left and right columns in Fig~\ref{Fig:psd} shows that  spectral curves for networks with inhibitory LTS and FS neurons are similar. Comparing the peak values indicated in the panels. we see that spectral power in the networks with FS neurons is slightly higher.

Remarkably, these power spectra, computed for single neurons, bear resemblance to experimentally obtained spectral curves. In the case of the voltage spectra (Figs.~\ref{Fig:psd} \textbf{B} and \ref{Fig:psd} \textbf{D}), the $1/f^{n}$ behavior is reported in experiments on up-down states with $n$ in the range 1 to 3 \citep{bedard2009,millman2010,baranauskas2011}. Furthermore, our results match the experimental observation that the spike-train power spectra have striking differences in comparison to the voltage-series power spectra \citep{bair1994} . 

In the spike-train power spectra (Figs.~\ref{Fig:psd} \textbf{A} and \ref{Fig:psd} \textbf{C}), there is no $1/f^{n}$ scaling. As the noise intensity $D$ is raised, the value related to the zeroth frequency bin of the spectra decreases. This  indicates that irregularity is becoming less apparent given that  
$\lim\limits _{f\to 0}\bar{S}(f)$ is related to the Fano factor which is a measure of irregularity \citep{middleton2003,pena2018}.

Regarding the $1/f^{n}$ scaling, observed both experimentally and theoretically \citep{beggs2003,kinouchi2006}, in our case we see that noise acts upon the scaling (cf. $n$ values in Fig~\ref{Fig:psd} \textbf{B,D}). It has been shown elsewhere \citep{baranauskas2011} that the shape of up-down transitions in the membrane potentials could be a determining factor for modulation of the $1/f^{n}$ scaling with $n=2$. Our observations provide support to this experimental evidence. At unbounded growth of $D$, transitions should vanish, and, as a consequence, $n$ decreases. 

Additionally, increase of noise shifts the peak values and peak frequencies
in both spike-train and voltage power spectra; compare the peak values in different subpanels. The existence of spectral differences where peaks becomes apparent or not is well known to be present in the cerebral cortex during different states such as slow wave sleep and wake \citep{buzsaki2006}.

We have seen that synaptic noise enforces alternation of collective states and influences durations of stay in each of them. 
Below, we explain how the dynamics of a single neuron, embedded in the synaptic noise setup, is reflected in the collective properties of activity, how the transitions are affected by the composition of the network, and how the picture changes at different levels of noise.

\subsection {Single neuron phase plane description of the synaptic noise setup}
\label{subsection_single}
A deeper understanding of the single neuron behavior in the synaptic noise setup can be gained from analysis of the course of its phase plane dynamics during the simulation. Setting the derivatives $\dot{v}$ and $\dot{u}$ in Eq.(\ref{eq:Izh-neuron}) to zero renders the nullclines of the voltage and the membrane recovery variable which we denote below as $\bar{u}$ and $u^*$, respectively.
\begin{eqnarray}
\begin{cases}
u & =\bar{u} = \alpha v^{2}+\beta v+\gamma + I(t), \\
u & = u^*= bv. 
\end{cases}
\label{eq:Izh-nullclines}
\end{eqnarray}
with $\bar{u}$ being a (time-dependent) quadratic parabola and $u^*$ a straight line. 
Synaptic noise enters this configuration implicitly, through its contribution to the current $I$.

Under the employed parameter values (see Table \ref{Tab::parameters} above) and $I=0$, the nullclines intersect in two points of the phase plane. These points correspond  to equlibria of the system; the left of them is stable: without input current, neuron exhibits no activity. When the instantaneous value of the current is increased, the nullcline $\bar{u}$ is shifted upwards on the phase plane, and the equilibria move towards each other.
At the value $I_{\rm sn}(t)=\displaystyle\frac{(\beta-b)^2}{4\alpha}-\gamma$ they merge and disappear in  a saddle-node bifurcation. 
Absence of equilibria is sufficient to ignite a spike:
the voltage grows until it reaches the threshold.
In fact, if the value of the parameter $b$ exceeds that
of the parameter $a$ (this holds for all four considered neuronal types), spiking starts at even weaker
current: at $I_{H}=\displaystyle\frac{(\beta-b)^2-(a-b)^2}{4\alpha}-\gamma$ the subcritical Andronov-Hopf bifurcation takes place, the equilibrium loses stability and the solutions spiral out from its vicinity towards the spiking threshold. Recall that the values of $\alpha,\,\beta,\,\gamma$ are common for all neuronal types (cf. Table~\ref{Tab::parameters}). Hence, the onset of spiking  at $I_{H}$ is dictated for each type of neuron by the pertaining $a$ and $b$ (the remaining parameters $c$ and $d$ characterize the reset and are irrelevant in this context: a neuron that has made it to the reset, is already in the spiking state).

Evolution of every individual neuron is governed by its instantaneous location on the phase plane with respect to the nullclines; its dynamics is affected not only by its own state, but by the time-dependent (due to external and synaptic currents) position of the nullcline $\bar{u}$. This allows us to see the collective dynamics from the local point of view of its individual participant; for it, the rest of the network is a background mechanism that moves the nullcline $\bar{u}$ upwards and downwards. 

Remarkably, this motion is not always negligible in comparison to dynamics
of the neuron on the phase plane: on arrival of synaptic input, the nullcline $\bar{u}$ 
is swiftly shifted in the vertical direction. Sometimes this leads to spectacular effects: 
a rapid fall of $\bar{u}$  may drag it across the instantaneous position of
the neuron on the plane and thereby halt and reverse the developing action potential. 
Such events, however, are seldom in a network like ours with its moderated connectivity, 
therefore most of the time the vertical displacements
of the nullcline $\bar{u}$ stay noticeably  slower than the motion of the neuron. 

\begin{figure*}[!ht]
\centering
\includegraphics[scale=0.27]{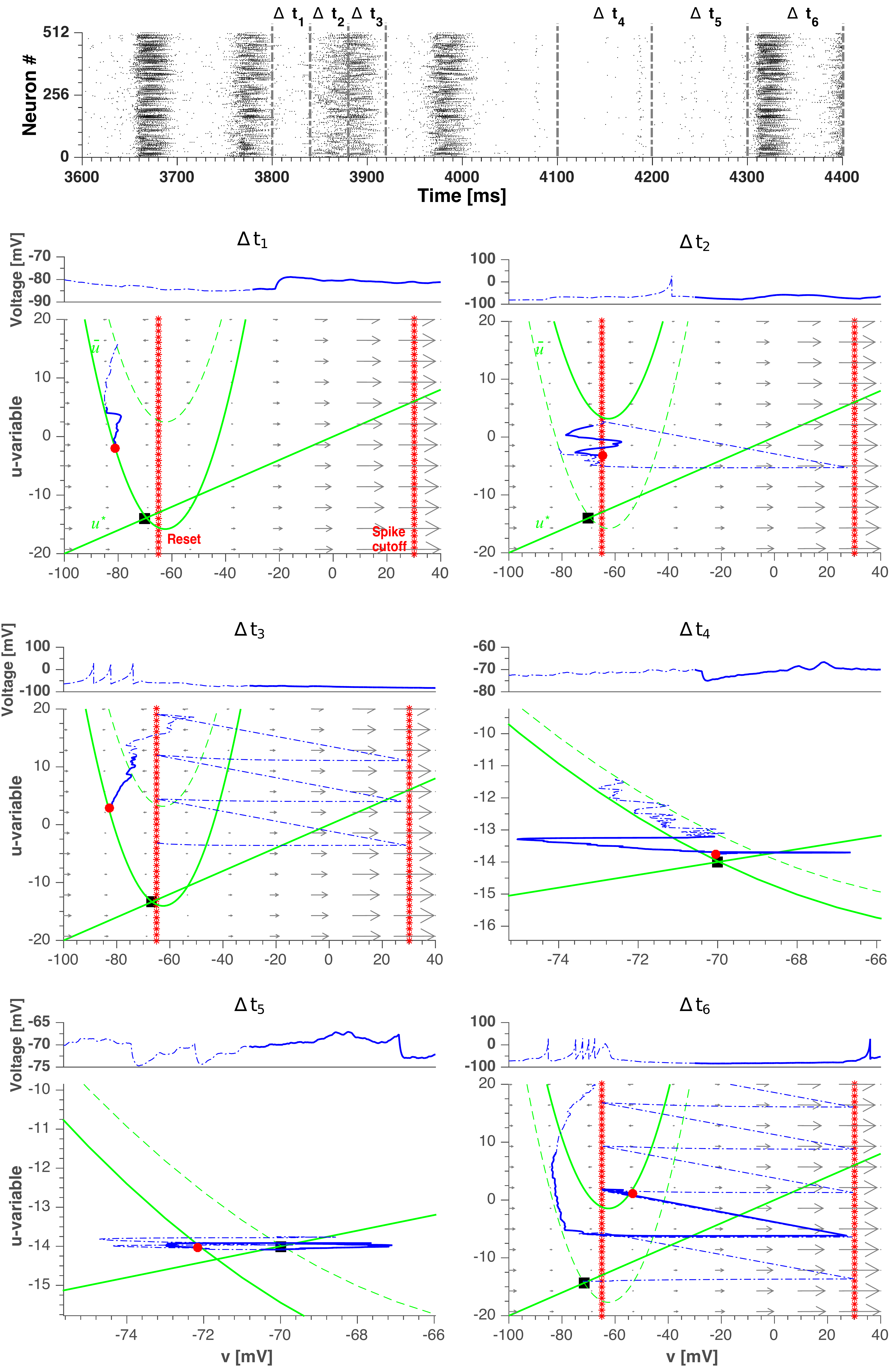}
\caption{\textbf{Single neuron phase plane depiction of a neuron that fires during the \textit{active} periods in the synaptic noise setup.} Upper panel: a zoom of the simulation from Fig~\ref{Fig:raster_sample}, split into 6 time intervals $\Delta t_i$. The first 3 intervals have a duration of 50 ms and the last 3 have a duration of 100 ms. 
Lower panels: voltage series and dynamics on the phase plane of neuron \# 240 in subsequent intervals $\Delta t_i$. Arrows indicate the vectors ($\dot{v},\dot{u}$); since $v$ is much faster than $u$, the vectors are nearly horizontal.
Blue dashed line: the first half of evolution  in a given $\Delta t_i$.  
Blue solid line: the last half of evolution  in a given $\Delta t_i$.
Red circle: location of the neuron at the end
of the time interval.
Black square: location of the state of rest with $v=v_{\rm rest}$ and $u=u_{\rm rest}$. 
Dotted red lines: reset value of voltage and spike cutoff. 
Green lines: Nullclines $\bar{u}$ and $u^*$, according to Eq.(\ref{eq:Izh-nullclines}). The location of the parabolic nullcline $\bar{u}$ is time-dependent; its position at the beginning (respectively, end) of $\Delta t_i$ is shown with dashed (respectively, solid) green line.}
\label{Fig:phase_plane}
\end{figure*}

\begin{figure*}[!ht]
\centering
\includegraphics[scale=0.4]{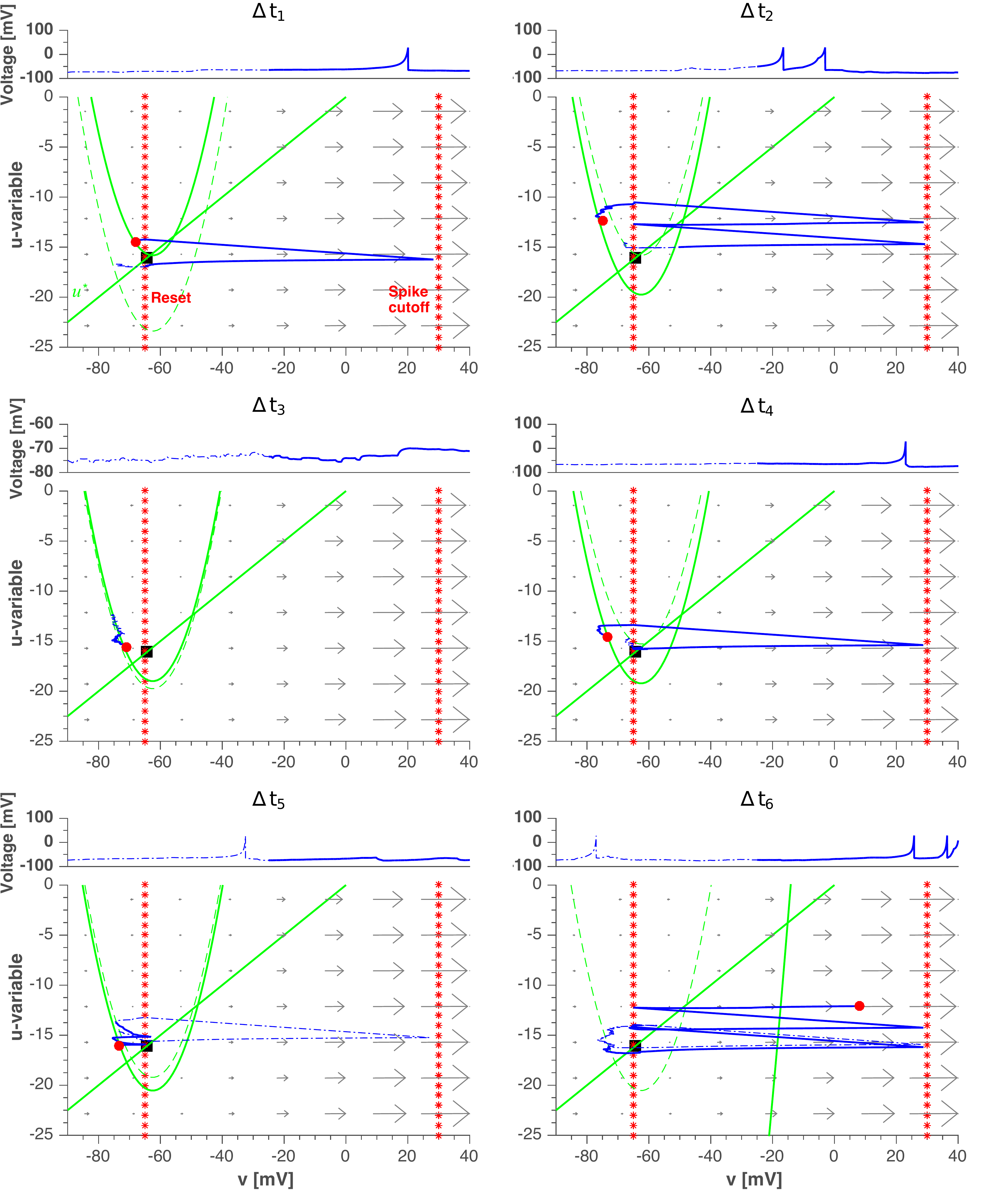}
\caption{\textbf{Single neuron phase plane depiction of a neuron that fires during \textit{all} periods in the synaptic noise setup.} Each panel contains voltage series and dynamics on the phase plane of neuron \# 69 for the same time range ($3800$--$4400$ ms) and the same six time intervals of $100$ ms as for the neuron in Fig~\ref{Fig:phase_plane}. Arrows in the plot indicate ($\dot{v},\dot{u}$).
Blue dashed line: the first 50 ms of evolution.  
Blue solid line: the last 50 ms of evolution.
Red circle: location of the neuron at the end
of the time interval.
Black square: location of the state of rest with $v=v_{\rm rest}$ and $u=u_{\rm rest}$. 
Dotted red lines: reset value of voltage and spike cutoff. 
Green lines: Nullclines $\bar{u}$ and $u^*$, according to Eq (\ref{eq:Izh-nullclines}). The location of the parabolic nullcline $\bar{u}$ is time-dependent; its position at the beginning (respectively, end) of $\Delta t_i$ is shown with dashed (respectively, solid) green line.}
\label{Fig:phase_plane_69}
\end{figure*}

With this local view in mind, we present in Fig~\ref{Fig:phase_plane} 
and Fig~\ref{Fig:phase_plane_69} the same simulation as in Fig~\ref{Fig:raster_sample} focusing on the individual dynamics of two representative neurons, arbitrarily picked among the populations of,
respectively, the neurons that fire only during the active periods and the neurons
that fire throughout all stages of evolution. As we will see, distinctions in the behavioral patterns can be traced down to the phase planes of the neurons.

We begin from the neuron \# 240 which fires only during the active periods, showing it in the time range between $3800$ ms and $4400$ ms. We split this range, which contains both active and quiescent states, into 6 smaller intervals $\Delta t_i$, each one of either $50$ or $100$ ms duration. The upper panel in Fig~\ref{Fig:phase_plane} shows the entire range and its breakdown into the set of $\Delta t_i$. The lower panels present for every $\Delta t_i$ the voltage series and the trajectory on the phase plane. Notably, in the hyperpolarized (down) state below reset, the neuron typically is close to the instantaneous location of $\bar{u}$, hence its motion is slow.

We summarize our observations as follows (for a clearer visualization of the moving trajectory we refer to the video in Online Resource 1):

\begin{itemize}
\item Interval $\Delta t_1$: in the beginning, the neuron has just ended its evolution in an up state and passes through a down state. There, the trajectory mostly stays inside the parabola of the voltage nullcline $\bar{u}$ below the reset value. Since the system is located above nullcline $u^*$ of the recovery variable, the latter decreases. The down state can be viewed as a period of relaxation where the voltage is hyperpolarized. The trajectory slowly moves towards the state of rest (marked as a black square in Fig~\ref{Fig:phase_plane}).
 
\item Interval $\Delta t_2$ : Before the trajectory arrives at the resting state, the neuron receives excitatory input from its presynaptic partners and the voltage nullcline $\bar{u}$ is shifted upwards, then the neuron resumes the up state and fires several times. The dynamics of $\Delta t_1 + \Delta t_2$ is largely repeated every $\approx 100$ ms.

\item Interval $\Delta t_3$:  Since most of the neurons are firing, their recovery variables are growing (recall that at every spike, $d$ is added to the value of the variable $u$). At very high $u$ the negative feedback to the voltage variable $v$ is so strong that the neuron is forced to stop firing and follows the same path as in $\Delta t_1$ (see \cite{tomov2016} for a description of this effect). The majority of the neurons in the network stops firing due to the same reason, and the network does not supply synaptic input, hence the conductances $G^{ex/in}$ relax. As a result, the nullcline $\bar{u}$ lowers and the neuron approaches the state of rest. 

\item Interval $\Delta t_4$: This is the middle of a quiescent period. The zoomed image shows how the neuron slowly moves towards the state of rest. The membrane recovery variable $u$ monotonically decays. Synaptic noise perturbs the trajectory, but falls short of initiating a new up state.

\item Interval $\Delta t_5$: the neuron crosses the nullcline $u^*$ of the recovery variable $u$. The latter does not decrease anymore while the voltage is fluctuating due to noisy synaptic input.

\item Interval $\Delta t_6$: finally the noise and/or arrival of inputs from presynaptic neurons are able to initiate a new active period.
\end{itemize}

The sequence of events in Fig~\ref{Fig:phase_plane} discloses a major role of the membrane recovery variable $u$ both in the transition from up state to down state and in the subsequent initiation of the new active phase by the noisy input. Because of high tonic firing during an up state, the total synaptic current into a neuron like 
\#240 is very intense and roughly constant (its fluctuation amplitude depends on the synaptic noise level). Hence, the nullcline $\bar{u}$ stays close to its highest position in the $u$-$v$ diagram while the neuron climbs towards it due to the increments received by its recovery variable $u$ after each spike. Finally, the neuron gets inside the parabolic nullcline $\bar{u}$, has its firing probability decreased and eventually stops firing. The fact that the whole network enters a down state when this happens suggests that most neurons behave like \#240, i.e. they dominate dynamics in the network. Excursion of the neuron to the left from the reset line while it is inside the parabola $\bar{u}$ is the mechanism responsible for the hyperpolarized voltages seen in the down states of oscillatory regimes both in the deterministic (cf. Fig~\ref{Fig:gex_gin_deterministic}) and noisy (cf. Fig~\ref{Fig:raster_sample}) setups. During a quiescent period, the nullcline $\bar{u}$ is dragged to the bottom of the diagram putting the neuron close to rest. This explains the absence of hyperpolarized voltages during quiescent periods (cf. Fig~\ref{Fig:raster_sample}). In this situation the neuron is also close to the nullcline $u^*$, so its eventual high jump to the region of the diagram below the nullcline $u^*$ makes the neuron fire again and a new active period begins.

The behavior of the neuron \#240 in Fig~\ref{Fig:phase_plane} somewhat mimics the overall behavior of the network: it is highly active during up states of active periods and silent during down states of active periods and quiescent periods. In the following, we will refer to neurons of this type as ``typical'' in the sense that they represent the behavior of the majority of the network nodes.

The firing pattern of typical neurons is contrasted  by the behavior displayed in Fig~\ref{Fig:phase_plane_69}. There, we show dynamics of the neuron \#69, chosen because of its atypical behavior: it fires at all stages: in the up and down states of the active period and during the quiescent period. Dynamical features of this neuron 
are complementary to the ones of the typical neuron in Fig~\ref{Fig:phase_plane}, and a combination of the views given by them offers a deeper understanding of the mechanisms responsible for the intermittent changes between active and quiescent states. 

A summary of our observations for the ``atypical'' neuron reads as follows (we refer the reader to the video in Online Resource 2 for a dynamical illustration of the effect):
\begin{itemize}
\item Interval $\Delta t_1$: contrary to the typical neuron, \# 69 starts its evolution with a low value of the recovery variable $u$. This indicates that during the previous up state the neuron did not fire much. The nullcline $\bar{u}$ also begins this time interval at a low position, meaning that it did not receive many increments. This suggests that the neuron is heavily inhibited when the network is at a high firing state, possibly being postsynaptic to a large pool of inhibitory neurons. Hence, it is more likely that the neuron emits spikes during down states: there it receives less inhibition from its presynaptic neurons, which, like the typical neuron in Fig~\ref{Fig:phase_plane}, are relaxing toward rest. Due to synaptic noise or eventual inputs from other similar neurons, the neuron \# 69 fires at a low rate during the down state.

\item Interval $\Delta t_2$: When the network enters the up state (second half of the time interval), the neuron is again strongly inhibited and emits fewer spikes than a typical neuron.

\item Interval $\Delta t_3$: The network up state continues and ends, whereas the neuron has a low probability of firing.

\item Interval $\Delta t_4$: This is the middle of the quiescent period. Note that by the end of the time interval the nullcline $\bar{u}$ moves down, indicating a net inhibitory input to the neuron (an early sign of the recovery of network activity which will come in the next time steps). Even weak synaptic noisy inputs can make it fire. Since the firing rate depends on the synaptic noise level, the duration of the quiescent period depends on it as well. 

\item Interval $\Delta t_5$: The situation is still as in the last time interval, but now we see a clear sign of the strong inhibition received by the neuron. After a spike in the first half of the time interval, when it is close to emitting a new spike, the neuron receives a strong inhibitory kick which hyperpolarizes its voltage and prevents the spike. The voltage grows again but another strong inhibitory impulse serves for the next setback. The inhibitory inputs come from neurons in the pool of presynaptic inhibitory neurons to \# 69, which are starting to ``wake up" on the eve of a new active period. As a consequence of the inhibitory inputs, the nullcline $\bar{u}$ moves further down.   

\item Interval $\Delta t_6$: The network enters the up state of an active period and most neurons are active again (like the typical \# 240 in Fig~\ref{Fig:phase_plane}). This makes \# 69 fire but because of the heavy inhibition, not at a high rate of the typical neuron. Evidence of the strong increase in the inhibitory input received by this neuron comes from the dramatic downward movement of the nullcline $\bar{u}$  out of the scale of the plot. 
\end{itemize}  

Excitatory neurons like the one in Fig~\ref{Fig:phase_plane_69}, which fire at low rates at all periods, will be called here ``quiet" neurons (elsewhere, in the context of the deterministic setup, we called them ``moderately active neurons" \citep{tomov2016}). Quiet neurons are fewer than typical neurons; for the network of Fig~\ref{Fig:raster_sample}, they, on the average, constitute about a quarter of the population.

The sequence of events depicted in Fig~\ref{Fig:phase_plane_69} underscores the importance of inhibition and synaptic noise in shaping the network activity during both down states and quiescent periods. Strongly inhibited during up states, the quiet neurons become disinhibited by the end of those states and serve as a source for
most of the spikes occurring during down states and quiescent periods.  
Thus, the firing pattern in the down states and quiescent periods is basically due to the recurrent excitatory synaptic connections among quiet neurons. 
The weak noise limit (cf. Fig~\ref{Fig:low_noise}) discloses the nature of the intrinsic activity pattern generated by the population of quiet neurons: it is highly asynchronous and non-oscillatory; remarkably, it is also weak. This confirms, on the one hand, that the population of quiet neurons is small, and explains, on the other hand, why the network activity during down states and quiescent periods is asynchronous and irregular. 

Due to the weakness of intrinsic activity of quiet neurons, the likelihood that their pool can trigger a high firing (up) state in the network is low and the synaptic noise level plays a pivotal role in controlling this likelihood. At low synaptic noise level, the weak activity of the quiet neurons can restore the up state when the network is at a down state, but this can be repeated generating a sequence of up-down oscillations only for a short transient time. An example can be seen in \nameref{S1_Fig}. After the transient the network enters a quiescent period: a persistent low activity regime characterized by asynchronous non-oscillatory activity. When the network is in a quiescent period, the activity of the quiet neurons is too weak to start a high firing state in the network; a certain minimal synaptic noise level is necessary to trigger this state. In the absence of this minimal synaptic noise level, the network activity remains in the quiescent regime as seen in the diagrams of Figs~\ref{Fig:diagram_rates} and~\ref{Fig:combined}. When the synaptic noise intensity increases above minimum level, the recurrent excitation amongst quiet neurons gets stronger, as well as the synaptic noise inputs to typical neurons, and the probability of the network exiting a quiescent period increases. 

The above discussion highlights a fundamental difference between down states and quiescent periods. In the weak synaptic noise regime, when the network activity is dictated by quiet neurons, their weak agitation is able to restore a high firing state in the network when the latter is in a down state but not when it is in a quiescent period. This phenomenon bears some similarity to the behavior observed previously by us in deterministic networks of two-dimensional nonlinear integrate-and-fire neurons in the absence of external inputs \citep{tomov2014,tomov2016}. There, the network state oscillates for a transient time between up and down states, before decaying to rest (cf. the behavior of the network in the deterministic setup in Sect. \ref{subsec:deterministic_setup}). The decay to rest always occurs when the state of the network in its high-dimensional deterministic phase space passes through a particular region of the phase space (a ``hole") which, when represented in the two-dimensional space of average voltage $\langle v \rangle$ and recovery $\langle u \rangle$ variables, overlaps with the region traversed by the network when it is in a down state \citep{tomov2016}. The analogy between down/rest state for the deterministic network without external input and down/quiescent state for the network in the synaptic noise setup suggests a further analogy between the hole in the high-dimensional phase space of the deterministic network and a hole in the high-dimensional phase space of the stochastic network. The difference is that when the network state in the stochastic high-dimensional phase space falls into its corresponding hole it escapes to a quiescent state instead of the resting state, and it can leave this quiescent state when the synaptic noise intensity is above a minimum level.

\begin{figure*}[!ht]
\includegraphics[scale=0.4]{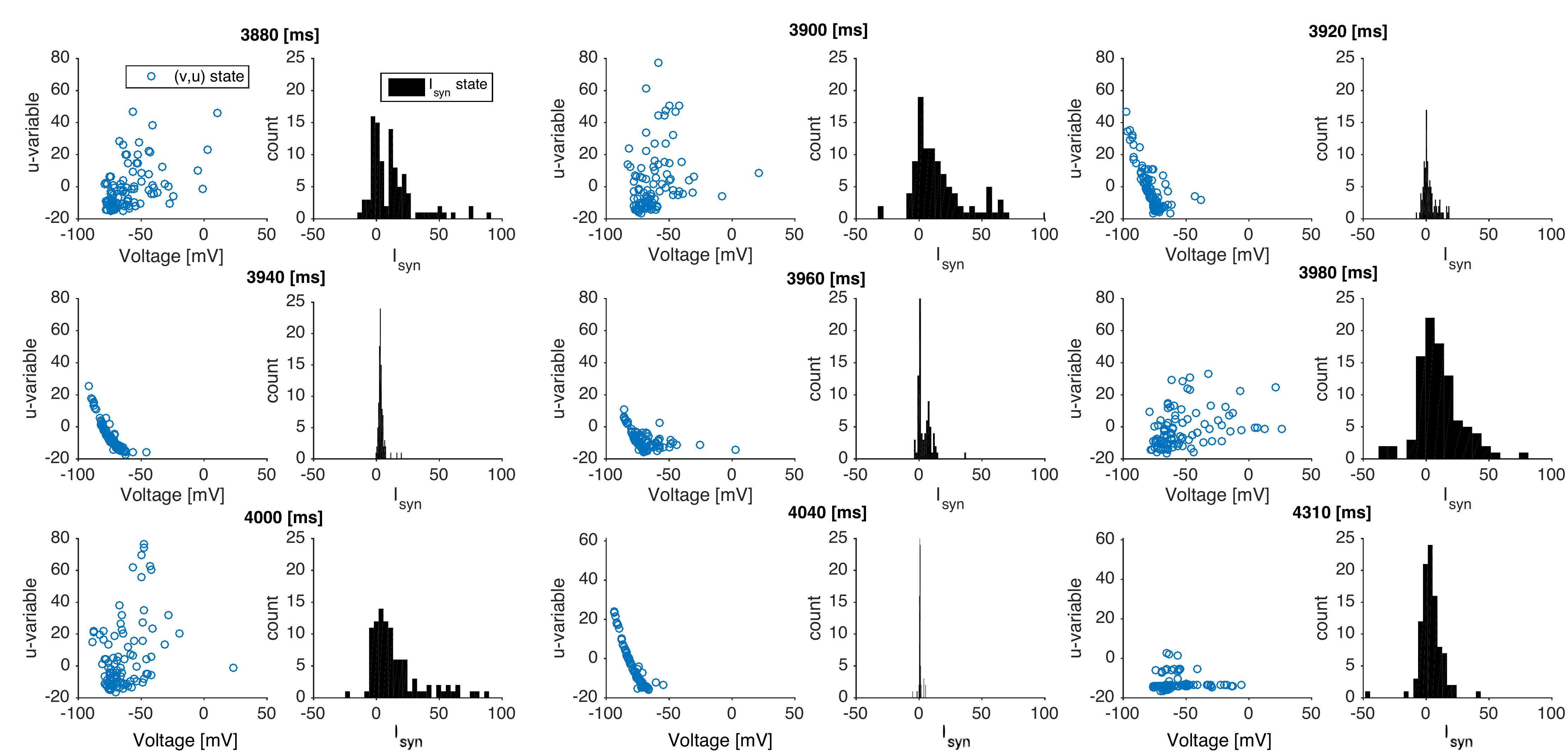}
\caption{\textbf{Distribution of the neuron variables and synaptic current at different moments of time.} Data in each panel come from 200 neurons pooled together from the same network simulation in Fig~\ref{Fig:raster_sample}. For different time instants, indicated atop every coupled subpanel, the figure presents scatter plots (left subpanel) of instantaneous ($v$,$u$) values, indicated as a blue circles, and histograms (right subpanel) of instantaneous $I_{syn}$ values. } 
\label{Fig:v_u_isyn}
\end{figure*}

To show that the recovery variable $u$ has a stronger impact on the cessation of activity than the inhibitory neurons, in Fig~\ref{Fig:v_u_isyn} we compare the effects of this variable and the synaptic currents $I_{syn}$ on the same neurons as in Fig~\ref{Fig:raster_sample}. In Fig~\ref{Fig:v_u_isyn} we present for selected time points both variables ($u,v$) for 200 neurons randomly picked from the network, and their total synaptic input $I_{syn}$. 

The first row in Fig~\ref{Fig:v_u_isyn} refers to an up-down transition: For $T=3880$ ms, which is the middle of the up state, some neurons have high values of $u$ (due to the constant increments the $u$ variable receives after each spike, cf. Eq.~\ref{eq:Izh-updates}) and consequently strong negative feedback. The consequence of this negative feedback is to hyperpolarize the neurons, which can be seen in the graphs for $T=3900$ and $3920$ ms where the voltages progressively move to the left of the graph. As to the $I_{syn}$ histograms, they are mostly dispersed around positive values (with a reduction in the dispersion as $T$ increases) indicating a low inhibitory activity. This confirms an earlier observation that the inhibitory neurons are not the main responsibles for the up-down transition  \citep{tomov2016}. The second row in Fig~\ref{Fig:v_u_isyn} refers to the down-up transition: for $T=3940$ ms, most neurons are hyperpolarized and the synaptic currents are sharply concentrated around zero, confirming that very few neurons (the quiet ones) are spiking, as shown in Fig~\ref{Fig:raster_sample}. As time increases, the distribution of neurons in the ($v,u$) plane becomes more disperse and the voltages $v$ move to depolarized values. This indicates that the neurons are free (without negative feedback) to spike again. Meanwhile, the distribution of synaptic currents widens-up and is dominantly excitatory (although there are some inhibitory currents). The third row in Fig~\ref{Fig:v_u_isyn} refers to the quiescent state: from $T=4000$ to $4040$ ms, the variable $u$ moves down and $v$ moves to hyperpolarized values. As observed in Fig~\ref{Fig:raster_sample} for the same condition, there are very few spikes. Only after about $300$ ms the voltages start to grow again and firing is re-started in a new active period.

\subsection {Influence of synaptic noise upon different states}

Having demonstrated in the previous section that synaptic noise affects different phases of activity, we now proceed to a quantitative description. We compute the average duration of active and quiescent periods in sufficiently long (we take the value of $6\times 10^5$ ms) trials. Mean duration is an important measure to characterize and model alternating states, e.g. in the course of transitions between brain rhythms \citep{lo2002,rubchinsky2013}. 

\begin{figure}[!ht]
\includegraphics[scale=0.3]{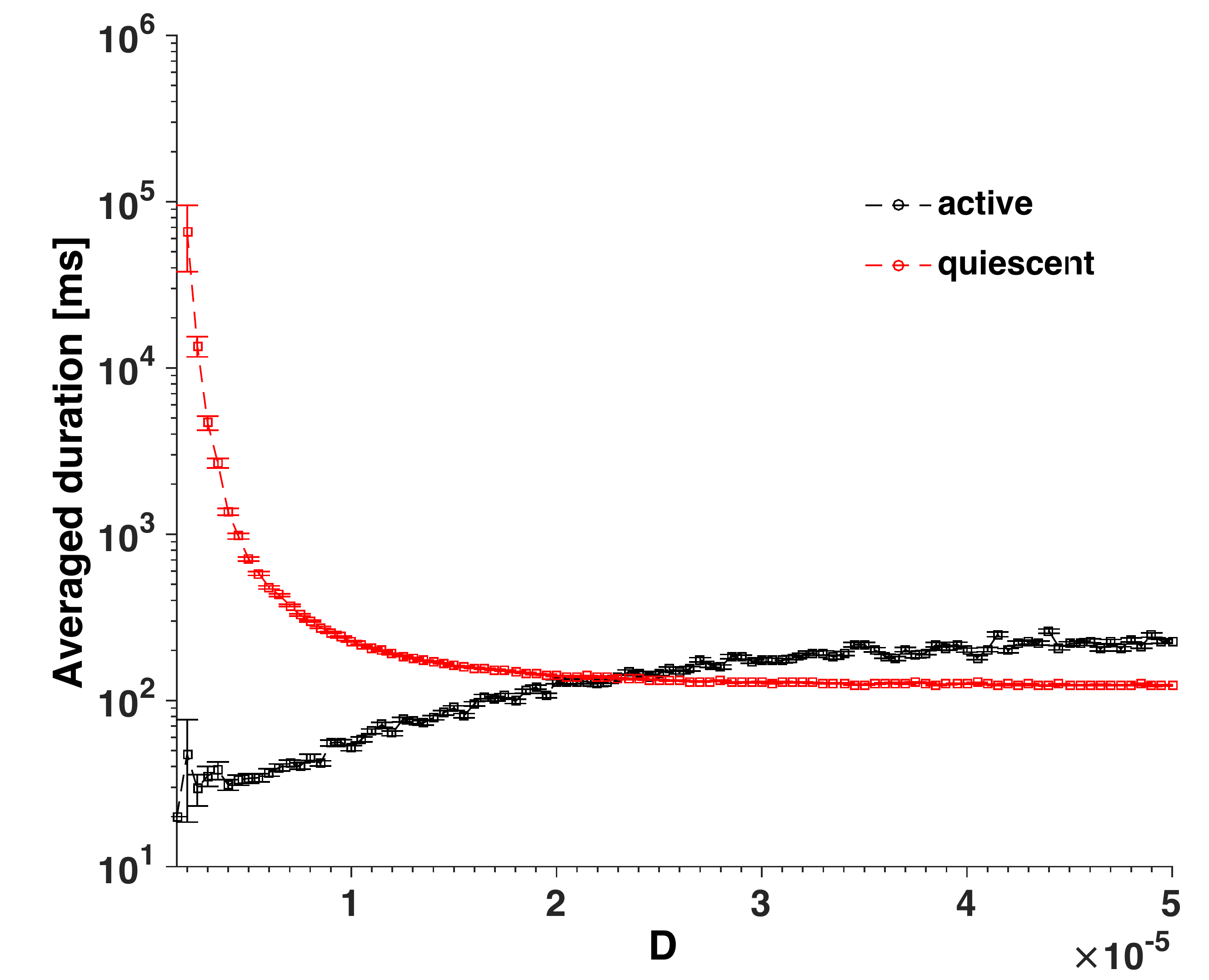}
\caption{\textbf{Synaptic noise intensity affects the mean duration of active and quiescent periods.} Curves show average durations of active and quiescent periods over the simulation of 10 min as a function of synaptic noise intensity. All inhibitory neurons are of LTS type. Synaptic noise intensity varies in the range 
$0.05\times 10^{-5}\leq D\leq 5\times 10^{-5}$ in discrete steps of size 
$\Delta D = 0.05\times 10^{-5}$.  Error bars: standard error.}
\label{Fig:resilienceFSLTS}
\end{figure}

Results of simulations confirm that the duration of stay in both active and quiescent periods is affected by the synaptic noise level (Fig~\ref{Fig:resilienceFSLTS}), but in a twofold way: the growth of noise intensity lengthens active periods and shortens the quiescent ones. This implies that synaptic noise influences transitions between the states. Remarkably, the average duration of stay in the quiescent state rapidly falls at the increase of small noise amplitude but seems to reach a certain saturation at moderate noise intensities. Apparently, the minimal time that the neurons need to organize a new collective activity is dynamically constrained by the network topology and deterministic characteristic times in the phase space: in the studied case it cannot be made lower than $\approx$ 80-100 ms.

\begin{figure*}[!ht]
\includegraphics[scale=0.35]{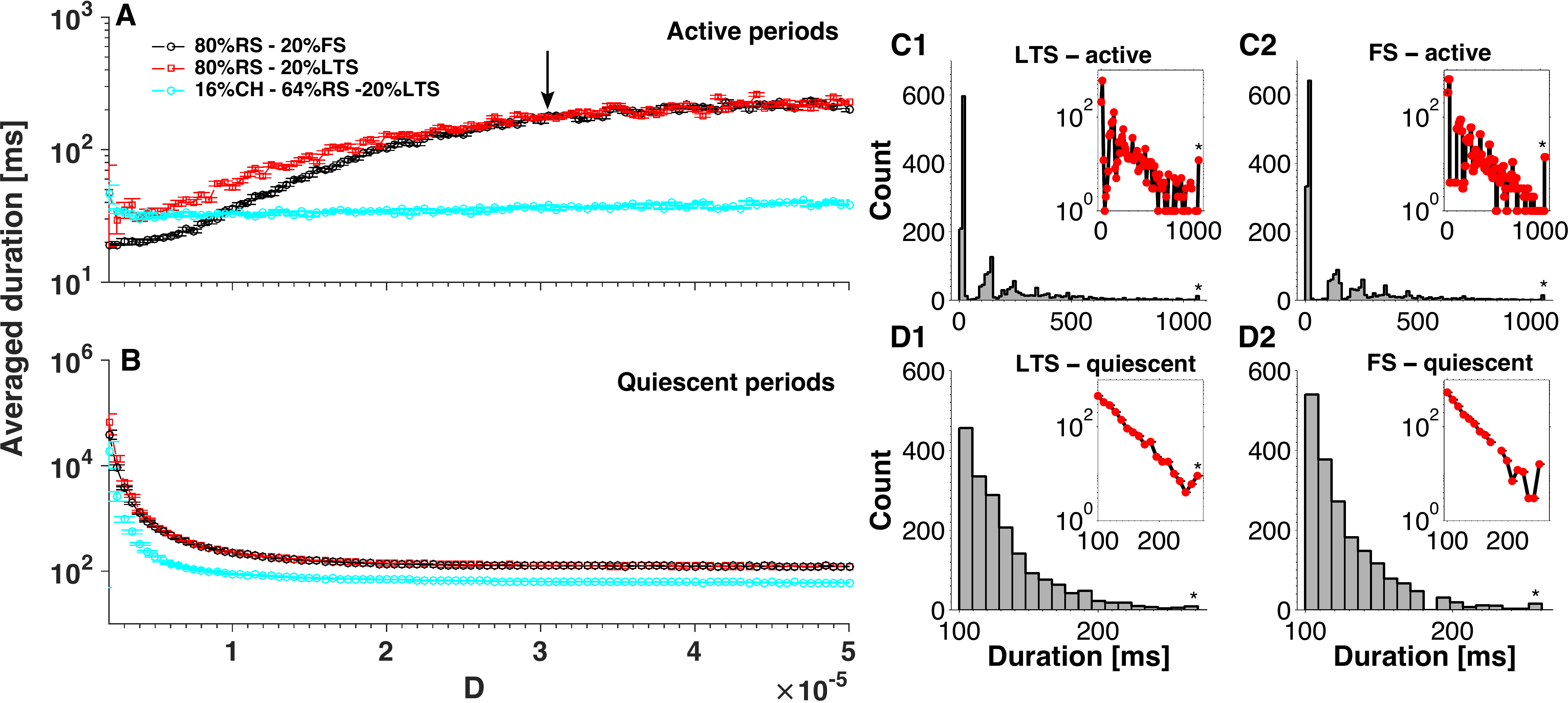}
\caption{\textbf{Network composition influences the average duration of stay in active and quiescent periods.} \textbf{A-B}: Dependence of duration on noise intensity. Legend in the plot indicates the network composition. Curves: average values over the simulation of 10 min. Error bars: standard error. \textbf{A}: active periods. \textbf{B}: quiescent periods. The value $D=3\times 10^{-5}$, denoted by the arrow, is used for calculation of histograms in panels \textbf{C1-2} and \textbf{D1-2}, characterizing distributions of stay duration in different periods. 
Stars at the end of the histograms are outliers. 
Insets show logarithmic representations of the ordinate.}
\label{Fig:resilience_lts_20p_ch}
\end{figure*}

Depending on the network composition, action of synaptic noise upon the average duration of active and quiescent periods can be weaker or stronger. Although the same common qualitative tendencies persist, quantitative aspects depend on the types of participating neurons as well as on proportions between them.
An exemplary comparison is shown in Fig~\ref{Fig:resilience_lts_20p_ch}. Simulations with two types of inhibitory neurons indicate that the LTS neurons, compared to the FS ones, seem to postpone the termination of the active period (top left panel): at low noise the duration of oscillatory activity is higher if LTS neurons are present. This implies that inhibitory neurons influence the transition from active to quiescent period. In contrast, the duration of the quiescent period (bottom right panel) displays no dependence on the type of inhibitory neuron: the corresponding curves in the plot nearly coincide. This indicates that the transition from quiescent to active period is regulated exclusively by excitatory neurons. Indeed, since inhibitory neurons cannot excite a network, every period of stay in the quiescent period should be interrupted by an excitatory neuron, or by a group of excitatory neurons. 

Introducing diversity among excitatory neurons, we observe certain quantitative changes as well. By replacing 20\% of RS neurons by CH neurons, we obtain a network built of 16$\%$CH, 64\%RS and 20\%LTS. This composition is much less sensitive to the action of synaptic noise. The tendency of growth of active periods under increase of noise is practically absent (see top left panel in Fig~\ref{Fig:resilience_lts_20p_ch}), and at all values of $D$ the average active period is shorter than the corresponding silent one. As for the latter, however, there is a systematic shift. Compared to the case when the whole excitatory population is of the RS type, in the mixture with CH neurons the mean duration of quiescent periods decreases to lower values, below $10^2$ ms. This decrease is a combination of synaptic noise- and network-related effects. A quiescent period ends whenever synaptic noise or some of the few quiet excitatory neurons which fire during the quiescent period (or both) drives across the firing threshold one of the majority of typical excitatory neurons which are at rest, provided that this neuron is able to activate its postsynaptic neighbors and initiate thereby a wave of activity. If the neighbors fail to fire, the quiescent period continues. The mean time required for the first neuron to fire is the same for the RS and the CH neuron (see Sect.\ref{sec::isolated}). However, the RS neuron issues just one isolated action potential, whereas the CH neuron generates a series of spikes, raising with each of them the conductances of its postsynaptic neighbors and creating thereby conditions for their activation and subsequent collective spiking. In this sense, a burst of a CH neuron has higher chances to initiate common activity than a spike of a RS neuron. Therefore, in a network with CH neurons the quiescent periods end earlier. This confirms our conjecture that excitatory neurons influence the length of quiescent periods.

Histograms of duration of stay in the active period in Figs~\ref{Fig:resilience_lts_20p_ch} \textbf{C1-2} show exponential distributions but are somehow fractured (cf. the logarithmic representations in the insets). Distributions of this kind have been reported previously \citep{duc2015, tomov2016}. In the  
former case the authors related cessation of activity to passages through a specific region in the phase space of their deterministic network (the ``hole" mentioned above), explaining thereby the quantization of cessation times. In our case, the behavior of the system is similar. Assuming the picture of a hole in the network phase space through which the network can escape from active to quiescent state, and a synaptic noise level high enough to allow multiple transitions from quiescent to active state, the quantization of active period durations can be explained keeping in mind that an active period is made of up-down cycles, each one with the same approximate period $T$. Since the escapes from active to quiescent state always occur at the end of an up-down cycle, the duration of an active state can only increase by integer multiples of $T$. The distributions of stay duration in the quiescent period, shown in Figs~\ref{Fig:resilience_lts_20p_ch} \textbf{D1-2}, possess exponential character as well, but without a fractured shape. This can be explained by the non-oscillatory nature of the quiescent periods.

It is important to note that the results concerning influence of variation of the synaptic noise intensity on mean duration of the different regimes can be translated to other features that in the end enhance the synaptic effect. For instance, if the noise intensity is kept constant but the network size is enlarged, the same effect is expected: the mean degree will increase and consequently the synaptic effect as well (cf. Eq.~\ref{Eq:synapse}). Our simulations with increased network size and connection probability have confirmed this effect; an example is presented in \nameref{S2_Fig} where we show that increasing the number of neurons to 5125 (5 times the standard network in this work) at constant $p$ creates a synchronous non-oscillatory activity type. In contrast, if $p$ is lowered, the network goes back to intermittent dynamics, although in \nameref{S2_Fig} the quiescent activity is rather rare (two periods are identified in a 60 seconds simulation) and for most of the time the network remains in the active state of up/down oscillations. 
Finally, a variation of the network size $N$ compensated by simultaneous change in the connectivity $p$, so that the mean degree $p N$ stays constant, keeps the synaptic effects (and, hence, the prevalent types of dynamics in the network) largely
unchanged.

\begin{figure}[!ht]
\includegraphics[scale=0.3]{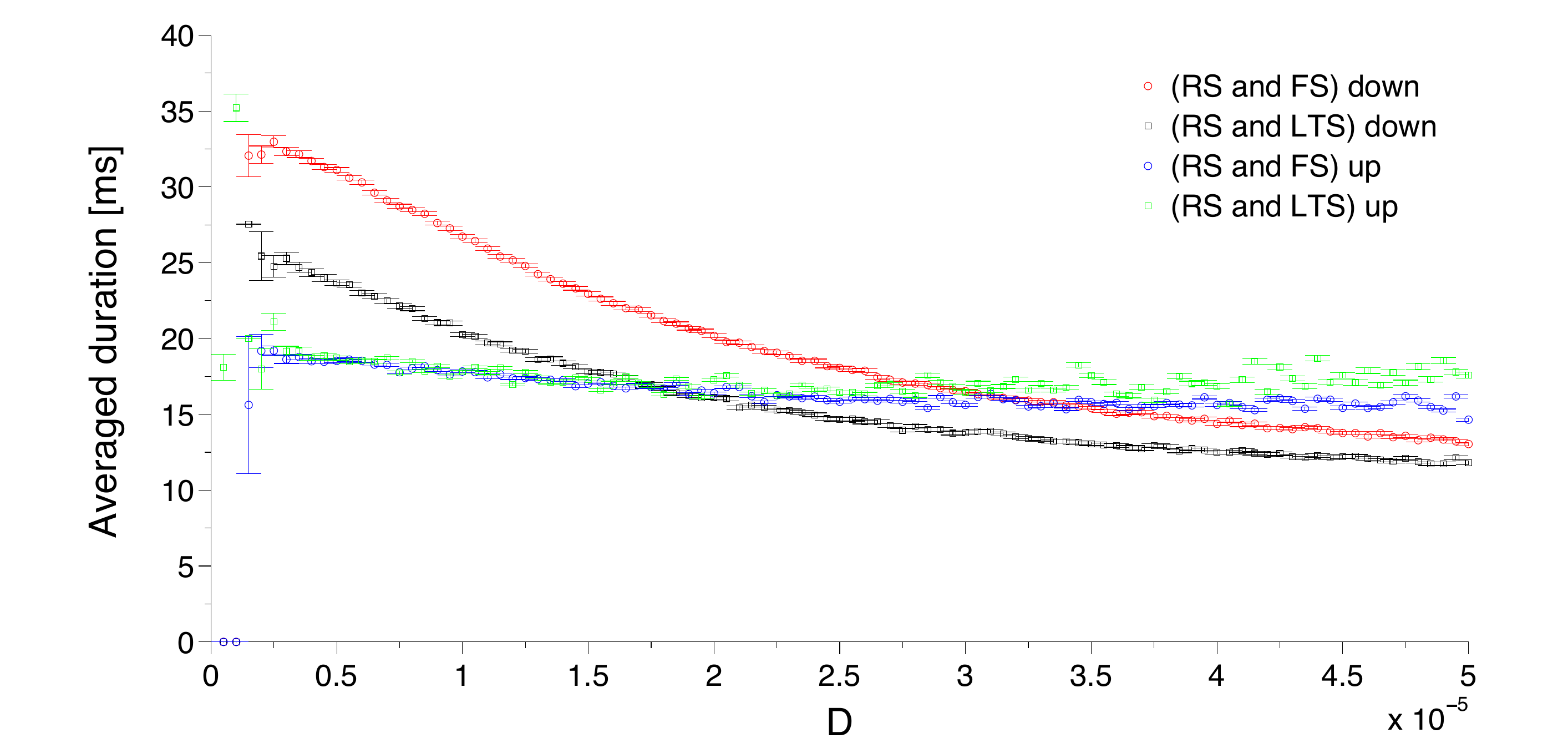}
\caption{\textbf{Average duration of stay in the up and down states as a function of noise intensity.} 
The network contains RS excitatory neurons and either FS or LTS inhibitory neurons.  
Curves: average values over the simulation of 10 min. Error bars: standard error.}
\label{Fig:resilience_up_down_FSLTS}
\end{figure}

Let us have a look at shorter timescales: what happens \textit{inside} the active periods? How does noise influence the collective up and down states? In Fig~\ref{Fig:resilience_up_down_FSLTS} we show dependence of average durations of stay in the up and down states on the noise intensity $D$, in the same range of $D$ as in Figs~\ref{Fig:resilienceFSLTS} and ~\ref{Fig:resilience_lts_20p_ch}.
Whereas the average stay in the down state gets shorter under the growth of noise, lifetime in the up state almost does not change.

The down state is the only state that is sensitive to the type of inhibitory neuron. There is a clear shift upwards (see red and black curves) if the LTS neurons are replaced with FS ones. This means that transitions from the down state happen faster in the presence of LTS neurons. The sensitivity of the down state can be related to the interpretation in~\cite{tomov2016} where the cessation of self-sustained oscillatory activity was assigned to passages through a small region of instability (the hole), located in the phase space close to the down state. In our current synaptic noise setup, the more noise, the higher the disturbance in the region of instability at the down state and the shorter its lifetime. 

\subsection {Comparison with other neuron models}
\label{subsect:comparison}
We expect the above results on intermittent transitions between active-quiescent states and the role of noise upon these transitions to stay qualitatively valid in networks based on other two-variable integrate-and-fire-type neuron models. To support this conjecture, below we apply the same procedure as in Fig~\ref{Fig:raster_sample} to the adaptive exponential integrate-and-fire (AdEx) model \citep{brette2005,gerstner2014}.

\begin{figure*}[!ht]
\includegraphics[scale=0.45]{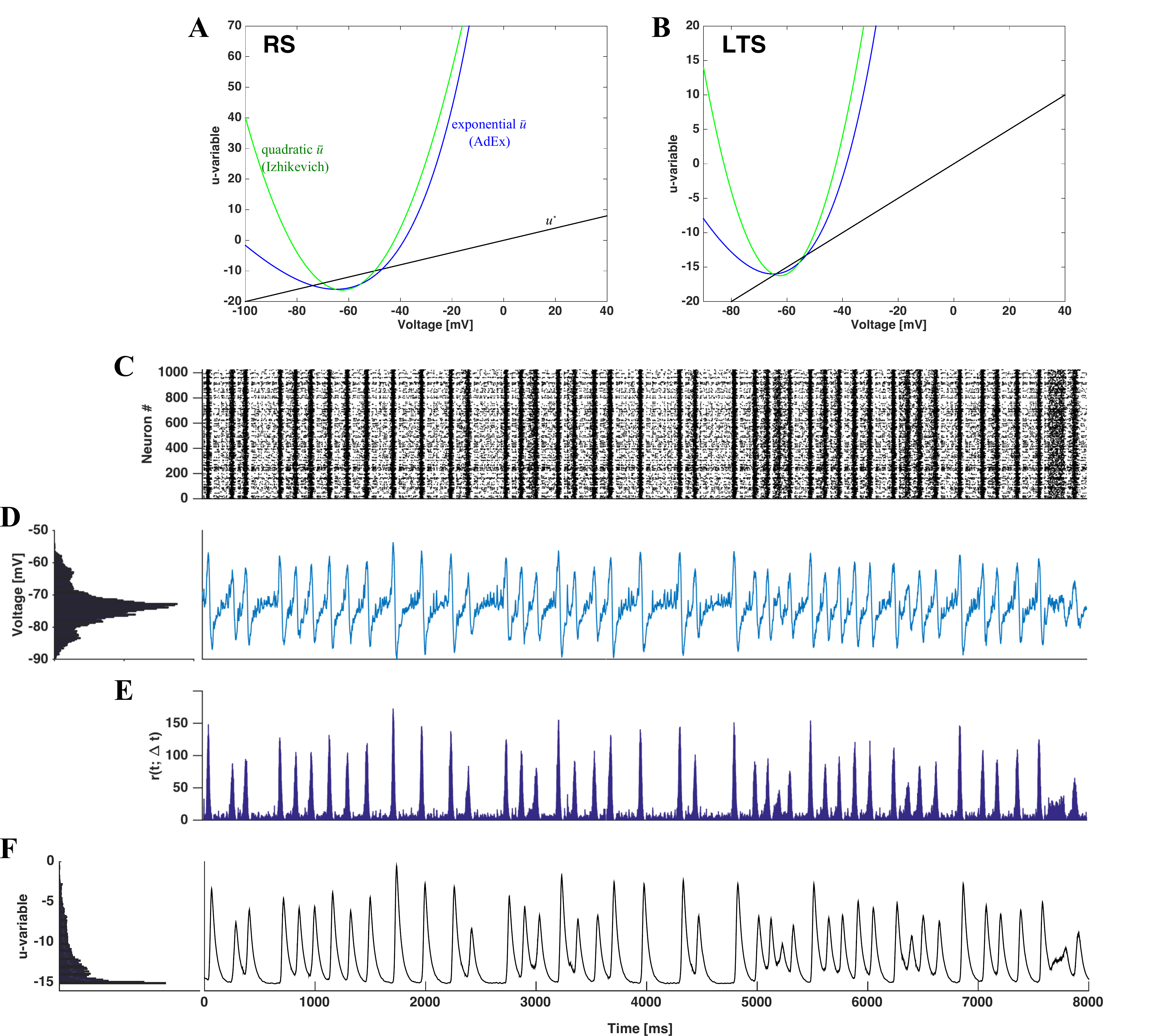}
\caption{\label{Fig:AdEx} {\bf Intermittent transitions between active and quiescent regimes in the presence of synaptic noise for network with AdEx neurons.} { The network is composed of two AdEx neuron types: the excitatory (RS) and the inhibitory (LTS), in the same proportion 4:1 as in the other networks of this work. Panels \textbf{A} and \textbf{B}: nullclines on the phase plane, drawn for the Izhikevich and the AdEx models for RS (\textbf{A}) and FS (\textbf{B}) neurons (see parameters in the text). 
Panel \textbf{C}: raster plot for a network simulation under synaptic noise  with $D=1\times 10^{-5}$. 
Panel \textbf{D}: averaged voltage histogram (left) and time course of averaged voltage over all network neurons (right).
Panel \textbf{E}: $r(t; \Delta t)$ extracted from all neurons in the network.
Panel \textbf{F}: Recovery variable $u$. Histogram (left) and course of $u(t)$ over all network neurons.
}
}
\end{figure*}

The AdEx model is a two-variable neuron that differs from the Izhikevich model by the equation for voltage: instead of a polynomial dependence on $v$, the AdEx features the exponential one. To run the AdEx network under the same conditions as the Izhikevich one without having to re-scale either the synaptic variables or the noise amplitude, we write the AdEx equations so that the input-frequency relationship and nullclines $\bar{u}$ and $u^*$ are similar to the ones from the Izhikevich model, leading to:
\begin{eqnarray}
\begin{cases}
\dot{v} & = -g_L(v - E_L) + g_L \Delta_T \exp{(\displaystyle\frac{v-v_T}{\Delta_T})} -46 - u + I(t) \\
\dot{u} & =a(b\,v-u), 
\end{cases}
\label{eq:AdEx-neuron}
\end{eqnarray}

\noindent where $\Delta_T=30$, $g_L$=1, $v_T=-65$, and $E_L=c$. The parameters ($a$,$b$,$c$,$d$) are the same as in the Izhikevich model. Along with Eq.~\ref{eq:AdEx-neuron}, the model includes the fire-and-reset rule given by Eq.~\ref{eq:Izh-updates}. A comparison of the Izhikevich nullclines to the AdEx nullclines is performed in Figs~\ref{Fig:AdEx} \textbf{A} and \textbf{B} where one can see that, for these chosen parameters and at the resting state ($I=0$), the fixed points for the two models are very close and the shape of the nullcline $\bar{u}$ is similar. 

In Figs~\ref{Fig:AdEx} \textbf{C-F} we see a qualitatively close behavior to  Figs~\ref{Fig:raster_sample} \textbf{A-D}: raster plots with $r(t; \Delta t)$ indicating that active (oscillatory) and quiescent (non-oscillatory) behaviors are switching sporadically, with voltages fluctuating among three different positions (hyperpolarized, resting, and depolarized), and recovery variable $u$ oscillating and featuring accumulation depending on the period. A noticeable difference however occurs when the average voltage in Fig~\ref{Fig:AdEx} \textbf{D} for the AdEx model is compared to the one in Fig~\ref{Fig:raster_sample} \textbf{B} for the Izhikevich model: for the AdEx model the voltage does not stay long enough in the hyperpolarized or depolarized states to create corresponding prominent peaks in the histogram (the peak for resting voltage is much more prominent). This difference is related to the integration of the AdEx neuron model, where the growth of voltage follows an exponential law, which is much faster than the quadratic one. This effect is reflected as well in the average voltage: the peaks and troughs are sharper than those for the Izhikevich neuron model. 

\section {Discussion}

Networks of LIF neurons have been extensively scrutinized in the literature to understand their properties under different conditions \citep{brunel2000dynamics,mattia2002,vogels2005,cessac2008,wang2011,litwin-Kumar2012,kriener2014,ostojic2014,potjans2014,yim2014,landau2016,jercog2017,tartaglia2017}. Much fewer works have been devoted to systematic investigations of networks of other spiking neuron models. Here we have studied networks of Izhikevich neurons in the presence of synaptic noise. We have found in these networks a rich variety of activity patterns, consisting of synchronous and asynchronous non-oscillatory states and oscillatory states with variable degrees of synchrony. Moreover, these networks exhibit intermittent noise-induced transitions between oscillatory and quiescent states. These transitions are irregular and affected by the synaptic noise level and the network composition. 

A systematic analysis of time series, plots of neuron spikes, firing rates, average voltage and membrane recovery variable, and power spectra revealed the characteristics of the oscillatory and quiescent states, similar to observed cortical states \citep{steriade2001,elboustani2007,greenberg2008,harris2011,sanchez-vives2017}: during oscillations the membrane voltages of the neurons fluctuate between hyperpolarized (down) and depolarized (up) states like in the so-called ``synchronized" states seen in \emph{in vivo} preparations and during slow-wave sleep and anesthesia; in the quiescent state neurons display very low and irregular spiking activity like in the so-called ``desynchronized" states seen in quiet rest. As far as we know, phenomena like oscillations between hyperpolarized and depolarized states, and noise-induced intermittent transitions between oscillatory and low activity regimes have not been reported in networks of LIF neurons.

By using the single neuron phase space representation of network dynamics combined with statistical assessments of duration of stay in the oscillatory and quiescent states, we were able to explain the roles played by synaptic noise and network composition in the durations of these states and the transitions between them. Besides, we also were able to explain the origin of the up and down oscillations and the asynchronous non-oscillatory nature of the quiescent states.

Up and down states, in which the average voltage of network neurons is, respectively, depolarized and hyperpolarized, occur during oscillatory (active) periods in the	 network. They can be understood in terms of the single neuron phase space in the same way as explained in the noiseless case \citep{tomov2016}. During an up state, when most of the neurons fire tonically, the parabolic-shaped voltage nullcline is kept in the upper part of the phase plane while the recovery variable moves steadily upwards due to neuronal firing. Eventually the neuron finds itself inside the area bounded by the voltage nullcline; it is forced to move to the hyperpolarized region of the phase plane and then downwards, relaxing towards rest. This corresponds to a down state. In the latter, the activity of the network is sustained by quiet neurons, which were inhibited during the up state and became disinhibited during the down state. In the course of time, firing of the quiet neurons is able to excite some of the relaxing post-active neurons; this triggers a new wave of excitation in the network, starting the next up state. This mechanism strongly depends on the recovery variable and its instantaneous increment (cf. Eq.(\ref{eq:Izh-updates})), which causes spike-dependent adaptation \citep{izhikevich2007}. Together, they constitute a sort of intrinsic negative feedback mechanism which decreases network excitability during the up state, as proposed by other authors in different contexts \citep{contreras1996,sanchezvives2000,bazhenov2002,compte2003,hill2005,holcman2006,parga2007,benita2012,chen2012,ghorbani2012,mattia2012,jercog2017,tartaglia2017,levenstein2018}.

The basic mechanism behind up and down oscillations is acting in both the deterministic and the synaptic noise setups. Thus, up-down oscillations are caused not by synaptic noise but by the intrinsic dynamics of the network. Disclosure of the same basic dynamical properties in the network with AdEx neurons (Sect. \ref{subsect:comparison}) allows to expect that this mechanism is common for networks populated by neurons with adaptation variables. In line with what has been pointed out elsewhere \citep{harris2011,mattia2012,jercog2017}, the up/down oscillations result from an interaction between recurrent synaptic connections and adaptation. 

Interestingly, the comparison between networks populated with Izhikevich and AdEx neurons indicates some differences between them: although the global dynamical behavior of the two networks is similar, the local voltage profile of their neurons is different (cf. Fig.~\ref{Fig:raster_sample} \textbf{B} and Fig.~\ref{Fig:AdEx} \textbf{D}). To the best of our knowledge, this is one of the first times in which the Izhikevich and AdEx neuron models are compared through their effects on the network.

The major difference between the deterministic and the synaptic noise setups is that in the deterministic case the oscillations are transient, while in presence of noise they become persistent. But the durations of the up and down phases and an up-down cycle are approximately the same, depending only on the  characteristics of network neurons. 

The up-down oscillations can be seen as a sort of default activity mode \citep{sanchez-vives2017} of the system  (at least in the region of the parameter space considered here). In the deterministic, noiseless, setup this activity eventually dies out, preceded, as we have shown, by the passage of the system through a specific region of its phase space we called a ``hole" \citep{tomov2016}. Through the hole, located close to the domain traversed by the system during a down phase, the system can escape the up-down oscillations and decay to rest. In the noiseless case the system sooner or later gets into the hole and the network activity dies out. In the synaptic noise setup, this hole-like region in the network's high-dimensional phase space still exists but because of the noise the system does not decay to rest when it passes through it; instead, the system is dragged to the quiescent state. 

As in the down state, in the quiescent state the network sustains activity, internally generated by quiet neurons via their recurrent synaptic connections and regulated by the synaptic noise level: it is weak for weak synaptic noise, and strong for strong synaptic noise. Being dictated by noise, activity during a quiescent period is asynchronous and irregular. Because of the passage through the hole the quiescent state has, in general, a longer duration than the down state. Hence, typical neurons which are relaxing in the hyperpolarized region of the single neuron phase space have time to decay to the phase space region around rest. This explains why during quiescent periods the average voltage is close to the resting voltage and is not hyperpolarized as in the down states. For weak synaptic noise, activity generated by the quiet neurons is insufficient to take the network out of the quiescent state: the system remains inactive. For moderate to high synaptic noise intensities, activity of quiet neurons gets stronger and even the neurons that are close to rest can fire, so eventually the global activity is reignited and an up state commences. 

The basic effect of the synaptic noise level is to increase/decrease the average duration of the quiescent periods. In other words, synaptic noise can act as a facilitator of transitions between quiescent and active states, and the intermittency between these states results from the stochastic nature of the neuronal firing during quiescent periods as well as from the irregularity of trajectory of the system in its high-dimensional phase space (that determines whether it will hit a hole). Once the system enters the hole, the duration of stay in the quiescent state depends on the noise intensity. For very low noise, the system stays in the quiescent state essentially forever, displaying only residual activity (see Fig~\ref{Fig:low_noise}). For moderate to high noise, the system eventually leaves the quiescent state and the up-down oscillations resume. The residence time in the quiescent state gets smaller as the synaptic noise intensity increases. For very strong noise the system may not even enter the hole because, in such a case, both typical and quiet neurons have high probability of firing at all moments. This explains the disappearance of quiescent periods in the high noise regime. For still higher levels of synaptic noise intensity, even the down periods disappear and the network features constant activity.  

Our study also indicates that inhibition affects transitions from active to quiescent periods and the duration of down states. Average stay in the down states is shorter when the inhibitory neurons of the network are of the LTS type than when they are of the FS type (cf. Fig~\ref{Fig:resilience_up_down_FSLTS}). This may be related to experimental evidence showing that inhibitory neurons control cortical oscillatory up and down states \citep{sanchez2010}. The authors of that study progressively blocked inhibitory cells during a spontaneous up state and showed that this blockage shortened the duration of up states and enlarged the duration of down states. Since the LTS neurons respond to noise faster, a replacement of all LTS neurons in the network by FS neurons can be viewed as a reduction of inhibition; thereby, the corresponding increase of the average duration of down states relates our observations in the model to the experimental evidence. Similar transitions from the up to the down states have been studied before \citep{holcman2006,xu2016}.

One of the objectives of our study was to check whether dynamics in the network of neurons with adaptation is sensitive to the composition of the network and the electrophysiological types of individual neurons. Generic qualitative features of dynamics, like  intermittent oscillations between active (up/down) and quiescent states, shape of power spectra, etc, turned out to be persistent for all neuronal subtypes as well as for their mixtures; on the individual level this can be traced back to the common shape of the nullclines. At the same time, we established that certain quantitative measures (like average durations) depend on the proportions of neuron types.

For noise, there are many ways to enter a neural network model \citep{faisal2008,longtin2013,destexhe2012,brochini2016,mcdonnell2016}. In this work we considered the variant in which it affects the synaptic variables. By doing so, we were able to study the effect of noise at the molecular level on the behavior of the system at the network level. Since noise at the synaptic level is related to fluctuations in the release of neurotransmitters and the amplitude of miniature postsynaptic currents \citep{rao2007,liu2010,tononi2014,kavalali2015}, which are phenomena at scales of magnitude much smaller than the scale of voltage changes, the weak noise intensities $D$ we considered here capture very small noisy events. Furthermore, because synaptic noise is filtered by the conductance variables, its effect upon neuronal voltages is akin to colored noise input, which is more biologically realistic than if noise were added, via e.g. Poisson processes, to neuron voltages directly. 

Our work captures mechanisms at different levels of neural processing with potential contribution to current endeavors to model multiscale brain mechanisms and their role on normal and pathological function \citep{mejias2016,neymotin2016,lytton2017,schwalger2017}. As an example, the synaptic noise-induced switches between periods of oscillatory and irregular activity might give support to fast formation and destruction of cell assemblies.

\begin{acknowledgements}
We are thankful to P. Tomov for stimulating discussions.
\end{acknowledgements}


\bibliographystyle{spbasic}      

%


\newpage

\section {Supporting information}

\paragraph*{Fig. S1 }
\label{S1_Fig}
{\bf Transient up-down oscillations in the weak synaptic noise setup.} Network composition: 16$\%$CH, 64\%RS and 20\%LTS neurons. Synaptic increments: $(g_{\rm ex},g_{\rm in}) = (0.15,1)$. Intensity of synaptic noise: $D=2.5\times10^{-6}$. The network received a brief external stimulus from the beginning of simulation until $t= 80$ ms. After this the network was left to evolve freely until the end of the simulation. Upper panel: raster plot showing the firing activity in the network for $t \geq 100$ ms after the end of external input. The raster plot is divided into 6 time intervals $\Delta t_i$ with duration 100 ms each. The network displays oscillatory behavior for a transient period followed by a quiescent state that lasts until the end of simulation.
Lower panels: voltage series and dynamics on the phase plane of an arbitrary selected neuron in subsequent intervals $\Delta t_i$. Arrows indicate ($\dot{v},\dot{u}$).
Blue dashed line: the first 50 ms of evolution.  
Blue solid line: the last 50 ms of evolution.
Red circle: location of the neuron at the end
of the time interval.
Black square: location of the state of rest with $v=v_{\rm rest}$ and $u=u_{\rm rest}$. 
Dotted red lines: reset value of voltage and spike cutoff. 
Green lines: Nullclines $\bar{u}$ and $u^*$, according to Eq (\ref{eq:Izh-nullclines}). The location of the parabolic  nullcline $\bar{u}$ is time-dependent; its position at the beginning (respectively, end) of $\Delta t_i$ is shown with dashed (respectively, solid) green line. 
\begin{figure*}[!ht]
\centering
\includegraphics[scale=0.3]{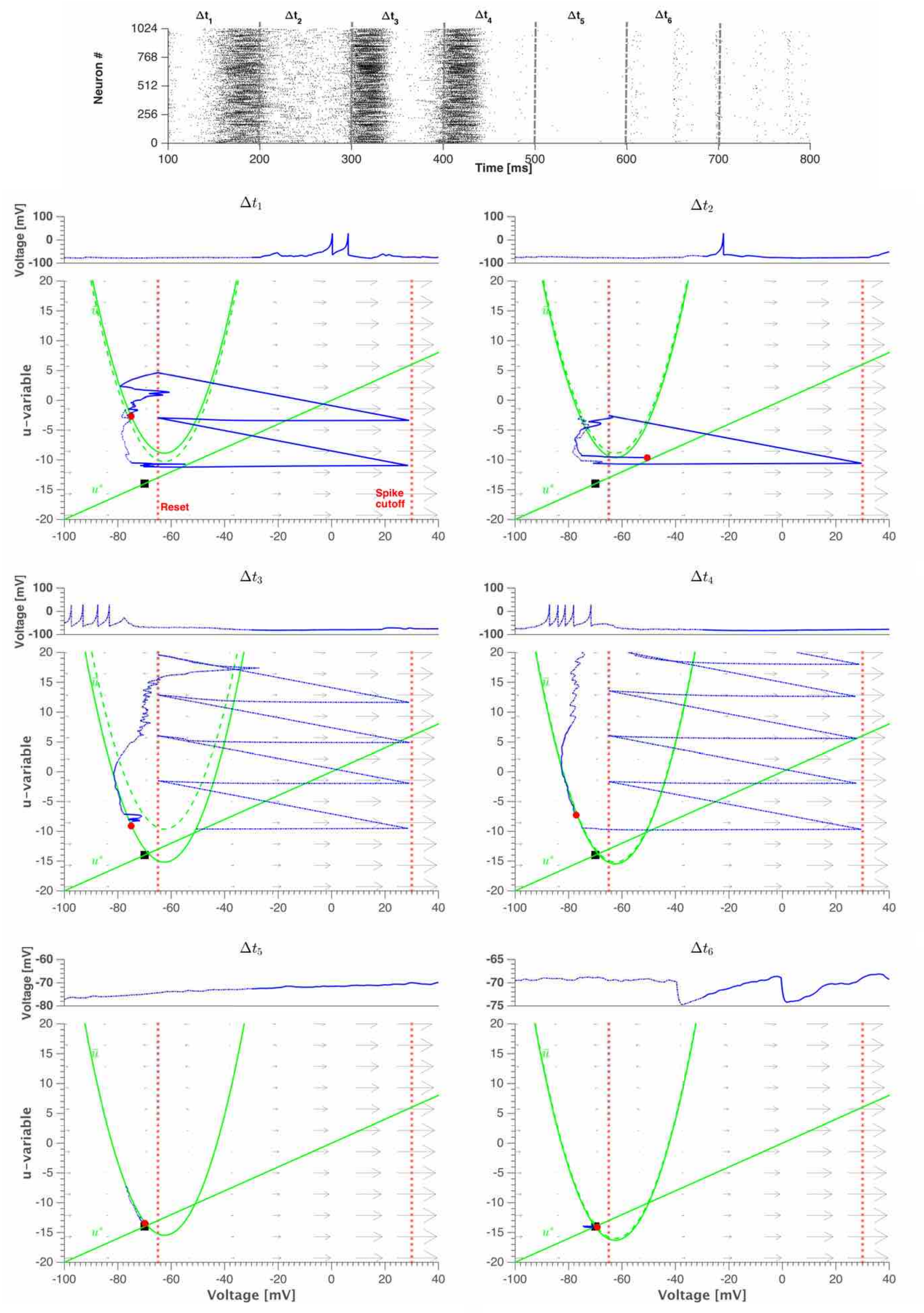}
\end{figure*}

\paragraph*{Fig. S2 }
\label{S2_Fig}
{\bf    Network size and connection probability dependence.} {In these simulations we used the same parameters as in Fig~\ref{Fig:raster_sample}, except the number of neurons and connection probability. The upper panels correspond to a network with 5125 neurons and connection probability $p=0.01$; the bottom ones concern the same number of neurons at $p=0.0005$. Different from Fig~\ref{Fig:raster_sample} one can see that enlarging the network size makes its dynamics to work in an asynchronous constant activity mode. In contrast to that, by lowering the connection probability, the intermittent activity with both quiescent and active periods present is recovered.}

\begin{figure*}[!ht]
\centering
\includegraphics[scale=0.4]{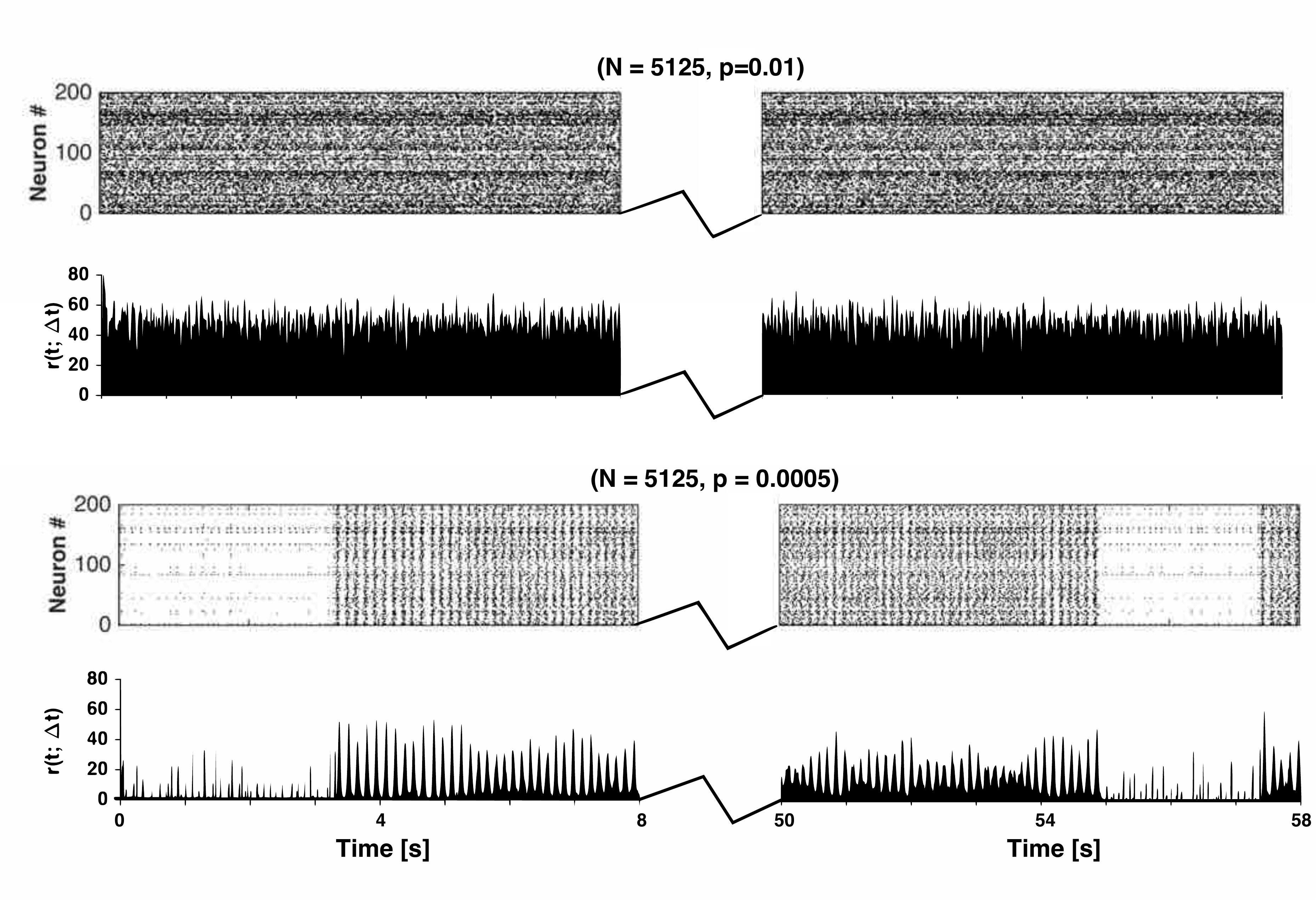}
\end{figure*}

\end{document}